\documentclass[12pt, a4paper]{article}
\usepackage[margin=1in]{geometry}
\usepackage{setspace}
\usepackage{graphicx} 
\usepackage{amssymb,indentfirst,verbatim,afterpage,caption,graphicx,subfigure, filecontents, bbm, enumerate, enumitem, tabularx, graphicx, fancyhdr,amsmath,latexsym,epsfig,amsthm,commath, epsfig,latexsym,float}
\usepackage[nointegrals]{wasysym}
\usepackage[table,xcdraw]{xcolor}
\usepackage[english]{babel}
\usepackage{natbib}
\usepackage{physics}
\usepackage{authblk}
\usepackage{multirow}
\usepackage[autostyle, english = american]{csquotes}
\usepackage[hidelinks=true]{hyperref}
\usepackage{tabulary,booktabs}

\MakeOuterQuote{"}
\usepackage{newunicodechar}
\newunicodechar{￥}{\textyen}

\usepackage{multirow}
\usepackage{arydshln}
\usepackage{tabularray}
\usepackage{stackengine}
\usepackage{threeparttable}

\usepackage{tabu,array}
\usepackage{amsfonts}
\usepackage{bbold}
\usepackage{mathtools}
\usepackage{booktabs}
\usepackage{rotating} 
\usepackage{graphicx}
\usepackage{soul} 
\usepackage{comment}
\usepackage{hyperref}

\usepackage{float}

\usepackage{xcolor}
\usepackage{newpxtext,newpxmath}

\usepackage{tikz}
\usetikzlibrary{shapes.misc}
\usepackage{subcaption} 

\usepackage{xcolor}
\definecolor{darkgreen}{rgb}{0.0, 0.5, 0.0}
\definecolor{darkyellow}{rgb}{0.8, 0.6, 0.0}

\newtheorem{hypothesis}{Hypothesis}
\newtheorem{result}{Result}

\newtheorem{obs}{Observation}

\begin{document}
\title{General Training and Worker Motivation: Experimental Evidence on Discretionary Effort
}

\author{Lawrence Choo, Senran Lin\thanks{Corresponding author.}, and Liangfo Zhao\thanks{Choo: University of Macau, Department of Economics. Faculty of Social Sciences Room 3034, Humanities and Social Sciences Building, University of Macau, E21B. Email: \href{mailto:lawrence.cy.choo@gmail.com}{lawrence.cy.choo@gmail.com}.
Lin: China Center For Behavioral Economics and Finance, Southwestern University of Finance and Economics, 555 Liutai Avenue, Wenjiang District, Chengdu, Sichuan, 611130, China. Email: \href{mailto:senranlin@outlook.com}{senranlin@outlook.com}. Zhao: China Center For Behavioral Economics and Finance, Southwestern University of Finance and Economics, 555 Liutai Avenue, Wenjiang District, Chengdu, Sichuan, 611130, China. Email:\href{mailto:liangfo.zhao@gmail.com}{liangfo.zhao@gmail.com}
We thank Guizeng Gao, Meng Xie, and Qing Zhang for their research assistance.\textbf{Data Availability Statement:} The data supporting the findings of this study are available in the Open Science Framework (OSF) repository at DOI \href{https://doi.org/10.17605/OSF.IO/TW7VH}{10.17605/OSF.IO/TW7VH}.This study was approved by the Institutional Review Board of the China Center for Behavioral Economics and Finance (CCBEF), Southwestern University of Finance and Economics, under protocol number 2022\_001. \textbf{Funding:} This research did not receive any specific grant from funding agencies in the public, commercial, or not-for-profit sectors.
}}
\date{\today}

\maketitle

\begin{abstract}
\noindent This study investigates the reaction of workers to employer-sponsored general training that provides skills useful not only in the incumbent employer but also in other firms in the industry. While previous research has focused primarily on workers' responses to wage renegotiation, our work extends this understanding by exploring an additional dimension---workers' discretionary effort beyond their job duties, which is not verifiable. We conduct a laboratory experiment to observe workers' responses in such an effort to different training intensities.
We find that workers generally increase their discretionary effort in response to general training, regardless of whether it is employer-sponsored or mandated. Moreover, the employer's intention behind offering training influences both effort and workers' renegotiation responses. Additionally, when workers can penalize employers, they do so, although higher employer-determined training intensities mitigate this behavior.
\end{abstract}


\onehalfspacing
\section{Introduction} \label{sec:intro}

\noindent The analysis of how labor acquires and is compensated for skills is fundamental to the study of labor economics. Firm-sponsored training is one of the most common and important approaches to enhancing worker skills and, thus, their productivity, which applies not only within the current firm but also across the industry. Among the different forms of training, \textit{general training}, which improves skills that are transferable across firms, has received considerable attention from researchers \citep{becker,bishop1996litreview,acemoglu1999beyondbecker,acemoglu1999structure,booth2002pays,manchester2012reimbursement}. 

Discussions of employer sponsorship of general training focus primarily on wage renegotiation after training (i.e., the improved job prospects for workers at the cost of employers). As argued by \cite{becker}, firms often lack incentives to sponsor general training due to the risk of post-training turnover and the higher wages needed to retain the trained workers---firms will eventually absorb the cost of training. According to Becker's \textit{wage-based argument}, these concerns prevent firms from benefiting from general training, making them reluctant to invest in it, despite its role in enhancing skills that are useful industry-wide and improving productivity.

Despite this theoretical concern, in practice, many employers invest in general training, even when they believe the skills are useful elsewhere \citep{loewenstein1999general}. Observations from industries such as financial services and retail (e.g., Starbucks' tuition reimbursement program) suggest that employers are willing to sponsor general training despite the risk of worker mobility. Empirical studies support these observations, documenting that a significant portion of training costs are covered by employers, even when employers acknowledge that such training, such as negotiation and presentation skills programs, enhances workers' value on the broader labor market \citep{smithhayton1999drives,manchester2012reimbursement}.

These findings challenge the predictions of classical models, which suggest that employers would avoid sponsoring general training without contractual guarantees of worker retention. This highlights the need to better understand workers' responses to general training, particularly when Becker's wage-based argument---and the broader literature following it---falls short.

Beyond affecting workers' post-training productivity and wages, general training can also influence their motivation in the workplace. Motivated workers are more willing to contribute to the firm's well-being, through both observable performance and unobservable efforts. We focus on \textit{discretionary effort}, which refers to non-contractible actions where workers ``go the extra mile'' beyond their job duties, such as maintaining the firm's reputation, referring potential employees, or fostering a positive workplace culture. Although often unobservable to employers, discretionary effort can have significant impact on employers' benefits.

Existing studies of general training focus primarily on its impact on wage renegotiation, and to a lesser extent, on observable performance \citep{sauermann2023effort}. However, non-contractable discretionary effort has received little attention. Unlike observable performance, which can affect workers' annual bonuses, thereby compounding monetary incentives with intrinsic responses to training, discretionary effort is non-verifiable and thus unlikely to be rewarded financially. Since it is privately known and not directly tied to monetary incentives, it remains an open question whether workers would put in such effort at all---purely out of intrinsic motivation---and whether workers' willingness to exert such effort depends on the employer's sponsorship of general training.

Against this backdrop, we conduct a laboratory experiment to investigate how workers respond to employer-sponsored general training, allowing us to directly measure discretionary effort. Specifically, we aim to answer the following question: \textit{How do workers respond to employer-sponsored general training, both in terms of discretionary effort and wage renegotiation?}

While discretionary effort is inherently difficult to observe or measure in field settings, laboratory experiments provide a controlled setting to investigate worker responses that are otherwise challenging to observe precisely \citep{charness2011lab}. Our experiment leverages this setting to directly study how it is affected by general training. Workers privately choose a level of discretionary effort for each training intensity, representing their hidden and unverifiable contribution to the employer's payoff. In this design, discretionary effort affects the probability of the employer receiving a high payoff, but the employer observes only the final payoff, not the chosen effort levels or probabilities.

Another advantage of our controlled experiment is the ability to make within-subject comparisons across different contingencies in training provision. Specifically, we study both workers' discretionary effort provision and their wage negotiation decisions in response to varying intensities of general training. In real-world settings, training provision varies due to factors such as personal ability, mobility costs, and the visibility of acquired skills \citep{acemoglu1998information,acemoglu1999beyondbecker,acemoglu1999structure,manchester2012reimbursement}. These factors are difficult to quantify, complicating comparisons of wage renegotiation outcomes between scenarios with and without training. Consequently, existing studies often focus on comparing the employer's gains against the costs of training. However, understanding workers' responses in the absence of training is equally important, as employers may invest in training to mitigate losses caused by its absence, even when productivity gains do not fully offset the expenditure on training. For example, the absence of training can lead to higher turnover intensity \citep{cappelli2004whypaycollege,pattie2006tuition,manchester2008effect}.

In our experiment, subjects are paired, with one taking the role of an employer and the other a worker. The employer and the worker interact under varying training intensities, which influence the worker's productivity and, consequently, her market value. We equalize the worker's market value with her productivity gain, ensuring that any increase in productivity is perfectly transferable and directly reflected in her market value. After training, the employer proposes a new wage, and the worker responds in two ways: through discretionary effort ($DE$), which reflects effort-based responses to training intensity, and through the \emph{minimum acceptable wage} ($MAW$), which captures wage renegotiation behavior. Furthermore, it is important to understand how $DE$ varies with training provision, particularly whether the worker's effort is driven by the employer's deliberate investment in training or merely by the training received (e.g., due to government-mandated training programs). To disentangle these effects, we implement two treatments: in the \textit{ENDO} treatment, training intensities are determined by the employer, while in the \textit{EXO} treatment, training intensities are assigned exogenously. Comparing these treatments sheds light on the role of intention in shaping effort-based responses.

We find that workers' discretionary effort ($DE$) increases in tandem with training intensities, confirming that $DE$ is a relevant behavioral response to training. This relationship holds even when training is imposed exogenously (\textit{EXO} treatment). However, at high training intensities, workers exert significantly more $DE$ when training is determined by employers. These findings suggest that workers respond positively to increased training, with their discretionary effort being particularly influenced by the employer's investment, underscoring the role of intention in motivating such effort.

Regarding wage renegotiation, we observe a significant reluctance to stay with employers in the absence of training, which diminishes at higher training intensities. This finding is consistent with prior empirical evidence that general training reduces turnover rates \citep{cappelli2004whypaycollege,pattie2006tuition,manchester2008effect}. At higher training intensities, workers adjust their acceptable wages below their market value to compensate employers, but these adjustments remain insufficient to fully offset training costs. This result aligns with the traditional view that general training is unprofitable for employers when evaluated solely on the basis of wage renegotiation. However, in the \textit{ENDO} treatment, when effort-based responses are accounted for, the complementary role of $DE$ in compensating training costs suggests that providing higher training intensities may be more profitable under our experimental conditions.

The \textit{EXO} and \textit{ENDO} treatments examine discretionary effort ($DE$) that benefits the firm. However, in many workplace settings, disgruntled workers may engage in discretionary effort that instead harms the firm's welfare, such as voluntarily posting negative feedback on an employee review platform and damaging the firm's reputation. Since workers voluntarily exert $DE$ in response to training, it is natural to investigate whether this effort extends to the negative domain when training is absent, as in the case of negative feedback.\footnote{\cite{charness2014dark} examine a related yet distinct form of counterproductive behavior, focusing on workers' harmful behavior toward co-workers rather than employers.} To examine this, we introduce two additional treatments, \textit{ENDO\_NEG} and \textit{EXO\_NEG}, which explore situations where a worker can exert \textit{counterproductive discretionary effort} ($CDE$) that directly reduces the employer’s benefits.

We find that workers consistently exert a positive level of $CDE$, even in the \textit{EXO\_NEG} treatment where training level is exogenously imposed, reflecting some degree of spite. However, between-treatment analysis shows that $CDE$ is reduced at higher training intensities when training is employer-determined rather than exogenously assigned. Furthermore, workers exert higher $CDE$ when the employer provides no training compared to when no training is assigned exogenously. With respect to the wage renegotiation behavior, it follows a similar pattern to that observed in the \textit{ENDO} and \textit{EXO} treatments and is not fundamentally altered by the reverse effect of effort-based responses.

We contribute to the training literature by demonstrating that workers' discretionary effort ($DE$) is indeed responsive to the employer's training intensity, even when it yields no direct financial benefit for the worker. This finding underscores the importance of discretionary effort in understanding the broader impacts of general training investments, as such effort can provide long-term benefits that extend beyond immediate productivity gains in various industries. For instance, in Information Technology, troubleshooting issues outside regular hours enhances service quality, fosters customer loyalty, and strengthens the employer's reputation. Discretionary effort also plays a critical role in fostering innovation. For example, in the education sector, employees spend time outside working hours developing creative teaching methods, which contribute to their institution's success.

Additionally, we broaden the scope of the training literature by introducing counterproductive discretionary effort ($CDE$). We find that $CDE$ is sensitive to the employer's intention in training investment. This finding suggests that potential losses resulting from the absence of training could be relevant to firms' investment in training, highlighting an important, yet less examined, dimension of training outcomes.

Most previous studies focus exclusively on post-training wage renegotiation, while some examine solely post-training performance, such as working hours \citep{sauermann2023effort}. Our study advances the training literature by integrating these two approaches, allowing the worker to respond through both a monetary channel (via the minimum acceptable wage, $MAW$) and an effort-based channel ($DE$), which is linked to a real-effort task. This combination can provide novel insights into how these responses interact and differ in shaping the broader impact of general training. Furthermore, our approach takes advantage of the unobservable nature of $DE$, ensuring that it cannot affect the employer's post-training offer and thus maintaining a clean separation between monetary and discretionary effort responses.

Our study also contributes to the literature on laboratory experiments examining reciprocal behavior by isolating intrinsic reciprocity through the use of unobservable responses. The gift-exchange game, introduced by \cite{fehr93giftexchange}, is a widely used leader-follower setting to examine how individuals reciprocate. For example, \cite{charness2004attribution} employ this game to disentangle the effect of the employer's intention on workers' responses. In a gift-exchange game, the leader makes an offer, and the follower responds with an observable action that incurs a personal cost but benefits the leader. The observability of the response introduces the potential influence of the \textit{signaling motivation}, where the follower's action is driven in part by an intention signaling her perception of the leader's offer---a phenomenon established in some leader-follower games \citep{xiaohouser2005emotion,xiaohouser2009avoiding}. In contrast, our setup incorporates an unobservable response---discretionary effort ($DE$). Since $DE$ remains unverifiable to the employer, our experiment isolates intrinsic reciprocity by eliminating the potential confounding effect of signaling motivation.

Furthermore, the gift-exchange game is predominantly used to study positive reciprocity, focusing on how the follower's response benefits the leader. In the workplace, however, negative reciprocity often plays a more significant and long-lasting role \citep{kube2006putting,kube2012currency,montizaan2016impact}. Our study contributes to this setting by introducing negative treatments in which the worker's discretionary effort penalizes, rather than rewards, the employer. This design allows us to investigate whether unobservable negative reciprocity channels in $CDE$ influence observable responses, such as the minimum acceptable wage ($MAW$), and to explore how these responses interact.

We outline the contextual background, experimental design, and procedures in Section \ref{sec:design}, including an alternative perspective on predictions with worker reciprocity in Section \ref{subsec_BE_model}. Section \ref{sec:results} presents our main results. Section \ref{sec:CDE} focuses on the two additional negative treatments, detailing their design and corresponding results. Finally, Section \ref{sec:conclusion} concludes the paper.

\section{Background and Experimental Design} \label{sec:design}

\subsection{Background}\label{sec: background}

\begin{figure}[htbp]
    \centering
    \includegraphics[width=\linewidth]{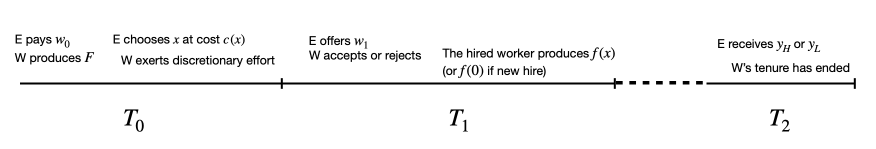}
    \caption{Timeline}
    \label{Figure_1}
\end{figure}

\noindent
We study the interactions between the employer (E) and worker (W) over three periods: $T_0$, $T_1$, and $T_2$ (see \autoref{Figure_1}).

In $T_0$, the employer pays the worker a fixed wage, $w_0$, corresponding to the market rate for an untrained worker. During this period, the worker generates output that contributes to the employer's revenue, with her productivity remaining at its initial level, denoted by $F$.

The employer also decides the worker's training level, denoted by $x \geq 0$, where $x = 0$ indicates no training. The employer fully bears the cost of this training, which is represented by a cost function $c(x)$, with $c(0) = 0$ and $c(x)$ increasing in $x$. A positive training level enhances the worker’s productivity to $f(x)$ in the subsequent period, $T_1$.

By the end of $T_0$, regardless of whether the worker has received training, she privately decides on the extent of discretionary effort (denoted by $DE$) to exert. This effort imposes a personal disutility on her but provides long-term benefits to the employer in $T_2$, benefits that are independent of her productivity in $T_0$ and $T_1$. The disutility associated with $DE$ need not be monetary (and not reflected in the monetary payoff); it may arise from the pain of executing effort. Examples of discretionary effort include referring potential employees, fostering a positive workplace culture, or enhancing the company's reputation. Importantly, $DE$ is unobservable and non-verifiable to the employer, and is therefore non-contractible.

In summary, the employer's $T_0$ monetary payoff is $m_{E}^{0}=F-w_0 -c(x)$, whilst the worker's $T_0$ monetary payoff is $m_{W}^{0}=w_0$.

At the beginning of $T_1$, the employer initiates the wage renegotiation process by offering a new wage, denoted by $w_1$. This wage cannot depend on the worker's discretionary effort, as it is unobservable to the employer. For simplicity, we frame this process as the employer making a \textit{take-it-or-leave-it} offer of $w_1$.

If the worker accepts the offer, she \textit{stays} with the employer: she receives $w_1$ and generates output $f(x)$ for the employer, where $x$ is the level of training provided in $T_0$, and $f(0) = F$. The productivity gain from training, $f(x) - f(0) \geq 0$, is referred to as the \emph{return of training}.  
In this case, the employer's $T_1$ monetary payoff is $m^{1}_{E,\text{stay}}=f(x)-w_1$, and the worker's is $m^{1}_{W,\text{stay}}=w_1$.

Alternatively, if the worker rejects the offer, she \textit{quits} the employer and receives the outside payoff $v(x)$. The employer is then assumed to hire a new untrained worker at the wage $w_0$, who produces output equal to $F$.\footnote{If the employer sets $x=0$ and offers $w_0$, he incurs no loss if the worker quits. This assumption ensures that no surplus unrelated to training is available for bargaining.} In this case, the employer's $T_1$ monetary payoff is $m^{1}_{E,\text{quit}}=F-w_0$, and the worker's is $m^{1}_{W,\text{quit}}=v(x)$.

A key assumption in our setup is that the training provides general skills, meaning that the return of training is \emph{perfectly transferable} --- for example, skills that are valuable to new employers or applicable in self-employment. As a result, the value of training in the labor market is:
\begin{equation}\label{eq:perfectly_transferrable}
    v(x)=w_0+f(x)-f(0),
\end{equation}
where the condition $v(x)-w_0=f(x)-f(0)$ indicates that the labor market fully appreciates the return of training.\footnote{We allow $f(0) > w_0$ to account for situations where the labor market does not fully reflect the worker's initial productivity. This is a weaker assumption than that of a perfectly competitive labor market, which would imply $f(0)=w_0$. Nevertheless, our assumption is sufficient to derive the non-profitability result in the classical model.}
In contrast to the previous literature (see, for instance, the discussion in \cite{acemoglu1999beyondbecker}), we do not impose additional assumptions, such as asymmetric information regarding its return, that would diminish the value of training (i.e., $v(x)-w_0< f(x)-f(0)$).

Period $T_2$ captures the long-term outcome for the employer, during which no further decisions are made by either party. This period extends beyond the worker's tenure, and thus, she receives no monetary payoff. The employer has a chance to receive a long-term benefit, denoted by either $y_H$ or $y_L$, where $y_H > y_L \geq 0$. Notably, this benefit does not \textit{directly} depend on the worker's production output. The probability of receiving $y_H$, denoted by $p(DE)$, increases with the level of discretionary effort ($DE$) the worker exerted after training in $T_0$. Thus, the employer's $T_2$ monetary payoff is $m_{E}^{2} = y_H$ with probability $p(DE)$, and $m_{E}^{2} = y_L$ otherwise. Since the worker does not participate in $T_2$, her monetary payoff is $m_{W}^{2} = 0$.

In summary, the employer's total monetary payoff across periods $T_0$, $T_1$, and $T_2$, is as follows: 
\begin{equation*}
    m_E=
    \begin{cases}
        F -w_0 - c(x) +f(x)-w_1 +y_H &\text{ if the worker stays and $m^2_E=y_H$,} \\
        F -w_0 - c(x) +f(x)-w_1 +y_L &\text{ if the worker stays and $m^2_E=y_L$,} \\
        F -w_0 - c(x) +F-w_0 +y_H &\text{ if the worker quits and $m^2_E=y_H$,} \\
        F -w_0 - c(x) +F-w_0 +y_L &\text{ if the worker quits and $m^2_E=y_L$.} 
    \end{cases}
\end{equation*}
For the worker, the total monetary payoff across all periods is:
\begin{equation*}
    m_W=
    \begin{cases}
        w_0+ w_1  &\text{ if the worker stays,}\\
        w_0+v(x) &\text{ if the worker quits.}
    \end{cases}
\end{equation*}

\subsection{Experimental Design}\label{subsec ExpDesign}
\noindent The experiment was conducted at the China Center for Behavioral Economics and Finance Experimental Economics Lab at Southwestern University of Finance and Economics in Chengdu, China, during November and December 2023.
The experiment included two treatments: \textit{ENDO} (60 subjects, 3 sessions) and \textit{EXO} (58 subjects, 3 sessions), which differ in how the worker's training level is determined. All participants were undergraduate students at the university. Each session lasted approximately $70$ minutes. The payoff exchange rate was fixed at $0.1$ CNY per point, with the average final payment being $51.43$ CNY. We will first describe the \textit{ENDO} treatment, followed by a discussion of how the \textit{EXO} treatment differs.

\begin{table}[htbp]
    \footnotesize 
    \centering
    \begin{tabu} to 1\textwidth {X[l] X[l] X[l]}
        \toprule
        \textbf{Treatment} & \textbf{Determinant of Training Level} & \textbf{Effect of DE on the Employer's Profit} \\
        \midrule 
        ENDO & Employer-determined & Increase the chance to obtain $y_H$ \\
        EXO & Exogenously assigned & Increase the chance to obtain $y_H$ \\
        ENDO\_NEG & Employer-determined & Decrease the chance to obtain $y_H$ \\
        EXO\_NEG & Exogenously assigned & Decrease the chance to obtain $y_H$ \\
        \bottomrule
        \bottomrule
    \end{tabu}
    \caption{Treatments Overview}
\end{table}

In the current section, we focus on the \textit{ENDO} and \textit{EXO} treatments, deferring the discussion of \textit{ENDO\_NEG} and \textit{EXO\_NEG} to the next section.

\subsubsection{ENDO Design}
\noindent The \textit{ENDO} treatment builds on the background outlined in the previous section. To isolate the reputation and sequential effects, the experiment is \textit{one-shot}.\footnote{Two practice rounds are conducted at the beginning, where subjects alternate roles as employer and worker, playing against a computer that makes random choices. After completing the current treatment, each subject enters a second treatment, \textit{EXO}, with the order counterbalanced (i.e., some subjects start with \textit{ENDO}, while others start with \textit{EXO}). However, only data from the first treatment are analyzed, so the experiment is treated as one-shot.} 
Meanwhile, the experiment seeks to frame a workplace environment where employers and workers typically build relationships, which is difficult to achieve in a one-shot setting. To facilitate this, we elicit subjects' attitudes on various topics (e.g., preferences for pets, opinions on carbon sustainability) and pair them based on similarity.\footnote{Our approach to matching subjects based on attitudinal proximity was inspired by \cite{Timm}.}

We set the initial conditions as $w_0=50$ and $F=100$. The employer begins by choosing a training level, $x$, from the set $\{0,1,2,3,4\}$. The employer bears the full cost of training, with $c(x)=20  x$, and the worker's $T_1$ productivity is given by $f(x) = F+100 x$. After deciding on the training level, the employer offers the worker a new wage, $w_1$.

The worker, after observing the training level $x$, privately chooses a level of discretionary effort ($DE$), ranging from $0$ to $12$, which corresponds to the number of tables in a real-effort task to be completed after the decision phase. Before making decisions, all subjects first complete a practice round of the task to familiarize themselves with its mechanics (see Subsubsection \ref{subsub:real-effort task}). 
To ensure that subjects realistically perceive the disutility or opportunity cost associated with discretionary effort, we employ this real-effort task (details of which are provided in Subsubsection \ref{subsub:real-effort task}) rather than a stylized design in which subjects merely select an effort level.
This method makes the experimental setting feel less artificial and allows real-life motivations influencing work behavior to play a role \citep{charness2018experimental, gill2011novel}. 
The chosen $DE$ increases the employer's likelihood of receiving the high-level long-term benefit, which is set as $y_H=800$, rather than the low-level one, set as $y_L=0$.\footnote{This likelihood of the employer receiving $y_H$ depends on the worker's chosen $DE$ at this stage, not her actual performance in the real-effort task. To discourage intentional overreporting, workers will receive bonuses if their actual performance exceeds their reported $DE$, with further details provided in Subsubection \ref{subsub:real-effort task}.} The likelihood is give by
\begin{equation*}
    p(DE)=0.2+0.05 DE \text{,}
\end{equation*}
where $p(DE)$ ranges from $0.2$ to $0.8$, depending on the worker's chosen $DE$ value in $\{0,\dots,12\}$.
The worker is also notified that while the employer learns whether $y_H$ or $y_L$ occurs, $DE$ and $p(DE)$ remain \emph{unobservable} to the employer, leaving the employer unable to distinguish if the worker chose $DE$ or if the observed outcome was merely due to chance.
Additionally, the worker submits a minimum acceptable wage ($MAW$). If the worker's $MAW$ is less than or equal to the offer from the employer, she accepts the offer and stays, receiving $w_1$ as her new wage; otherwise, she rejects the offer and quits, obtaining her outside payoff $v(x)=w_0+f(x)-f(0)=50+100x$.

In our experimental, we employ the \textit{strategy method} \citep{brandts2011strategy}, where subjects act as both the employer and the worker before their roles are assigned. This method allows subjects to consider the perspectives of both roles and enables us to compare within-subject responses to varying training levels. Specifically, as workers, they submit $MAW$ and $DE$ for each possible training level, $x \in \{0,1,2,3,4\}$. Once roles are assigned, the $MAW$ and $DE$ corresponding to the employers' chosen $x$ are implemented. The realized $DE$ then determines the likelihood of the employer receiving of receiving the high long-term benefit ($y_H$).

The parameterized total payoffs are as follows:

For the employer,
\begin{equation}\label{eq_E_total_payoff}
    m_E=
    \begin{cases}
        150-w_1+ (100-20) x+800 &\text{ if the worker stays and $m^2_E=y_H$,} \\
        150-w_1+(100-20)  x  &\text{ if the worker stays and $m^2_E=y_L$,} \\
        100-w_1-20 x +800 &\text{ if the worker quits and $m^2_E=y_H$,} \\
        100-w_1-20 x &\text{ if the worker quits and $m^2_E=y_L$.} 
    \end{cases}
\end{equation}

For the worker,
\begin{equation}\label{eq_W_total_payoff}
    m_W=
    \begin{cases}
        50 + w_1  &\text{ if the worker stays,}\\
        100 + 100 x &\text{ if the worker quits.}
    \end{cases}
\end{equation}

\subsubsection{EXO Design}
\noindent Having described the \textit{ENDO} treatment, we now introduce the \textit{EXO} treatment, which shares the same design but differs in one key aspect: the training level is assigned exogenously, determined randomly with equal probability for each level of $x \in \{0,1,2,3,4\}$. After observing the assigned training level, the employer offers a new wage, $w_1$. This variation, compared with the \textit{ENDO} treatment, enables us to isolate the effect of the employer's intent in sponsoring training on the worker's behavior.

\subsubsection{Real-effort Task}\label{subsub:real-effort task}
\noindent After subjects make their decisions, those assigned as workers proceed to the real-effort task. This task enables subjects to experience the disutility of performing discretionary effort.

\begin{figure}[H]
    \centering
    \includegraphics[width=\linewidth]{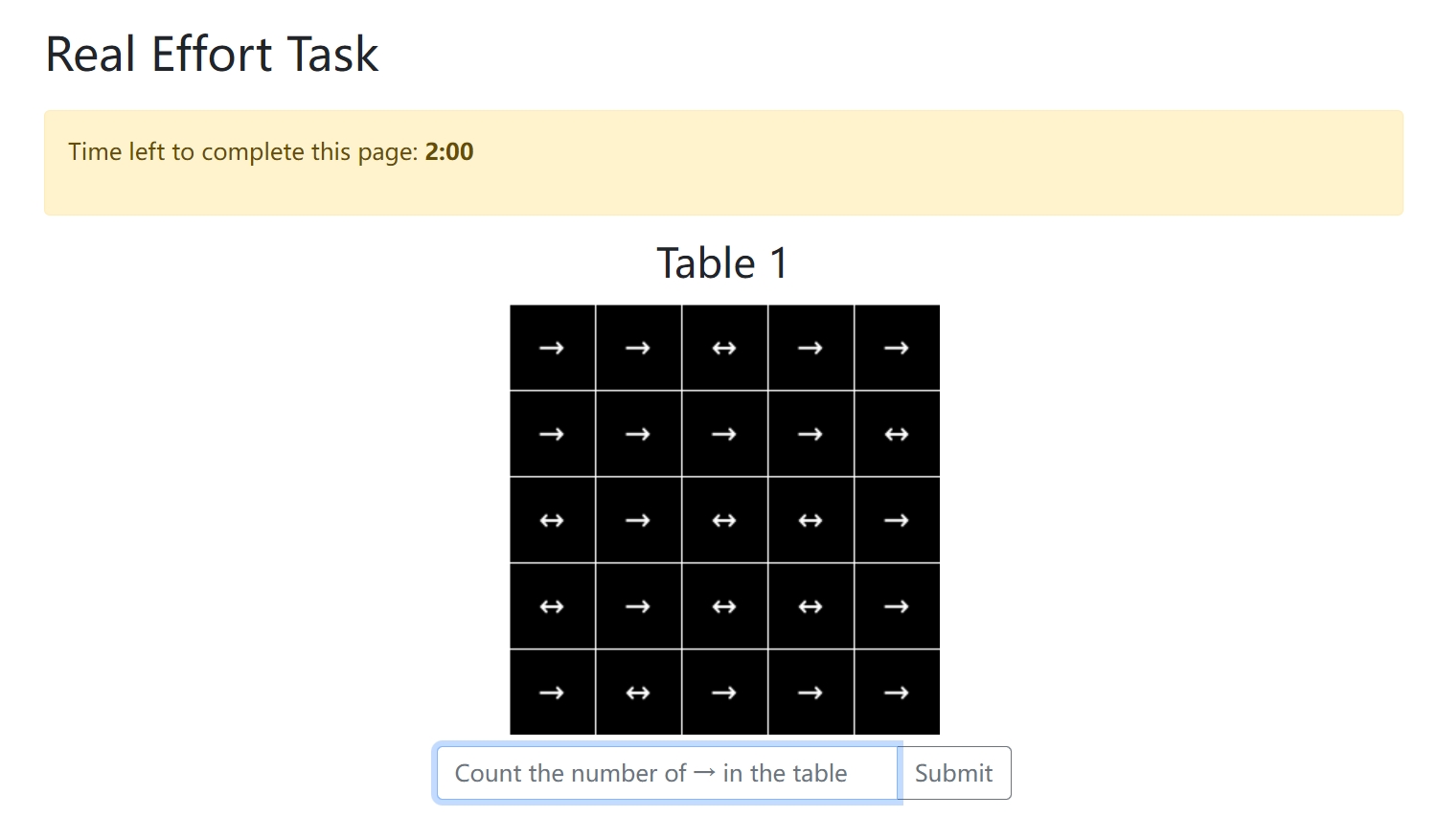}
    \caption{Screenshot of the Real-Effort Task} 
    \label{Figure_2}
\end{figure}

The real-effort task involves counting right arrows in a table containing both right and horizontal double arrows, with a two-second delay for incorrect entries before retrying (see \autoref{Figure_2}).

In the experiment, an employer's payoffs depend on his worker's chosen $DE$ in the decision phase, rather than her actual performance, ensuring that $DE$ reflects the worker's intent rather than her ability to perform the task.\footnote{An alternative design could tie the employer's likelihood of receiving $y_H$ to the worker's actual performance. However, this approach poses challenges due to ability heterogeneity and requires the worker to assess her ability to meet the chosen $DE$. This shift from intrinsic motivation to self-assessment could distort $DE$ as a measure of the worker's intent to reward her employer and reduce its comparability across subjects.}

However, this approach carries the risk that the worker may complete fewer tables than her chosen $DE$, since actual performance does not influence the employer's payoffs, and there is no direct penalty for underperformance. In this case, the worker benefits the employer without experiencing the disutility of performing the real-effort task. 

To mitigate this, we introduce additional incentives: a worker receives $2$ CNY for each table completed beyond her reported $DE$, incentivizing her to exert her best effort during the task.\footnote{Our empirical results, discussed in Section \ref{sec:results}, indicate that the issue of overreporting $DE$ rarely occurs.} 

At the same time, the bonus imposes an opportunity cost when a worker allocates $DE$ for her employer's benefit, reflecting the trade-offs workers face when exerting discretionary effort in real-world settings. Specifically, if a worker chooses zero $DE$, she can collect bonuses from all completed tables.

\subsubsection*{Post-experiment Survey}  
\noindent At the end of the session, subjects complete a non-incentivized post-experiment survey. The survey gathers data on social value orientation \citep{murphy2011measuring}, reciprocity tendencies \citep{perugini2003personal, dohmen2009surveywage, montizaan2016impact}, and demographic characteristics.

\subsection{Hypotheses}
\noindent Our hypotheses are based on classical models of worker behavior, such as those formalized by \cite{becker}, which assume workers are self-interested and aim to maximize their income. In this framework, workers are predicted to accept a new wage only if it meets or exceeds their market value $v(x) = w_0 + f(x) - f(0)$. Because $f(x) - f(0)$ captures the full return of training, the employer must compensate the worker entirely for this gain through the new wage. Consequently, the employer derives no benefit from training, making it unprofitable to sponsor any positive level of general training.

In the experiment, we collect the worker's minimum acceptable wages for each training level, denoted as $MAW(x)$, influenced by two factors: the outside option payoff, $v(x)$, and the worker's motivation to respond to the training. To capture the latter, we derive the \textbf{relative wage gap} ($RWG(x)$), which measures the percentage difference between the market wage, $v(x)$, and the worker's $MAW(x)$:
\begin{equation*}
   RWG(x)= \frac{v(x)-MAW(x)}{v(x)}.
\end{equation*}
Here, the numerator, $v(x)-MAW(x)$ represents the \emph{wage gap}. A positive wage gap (hence, $RWG(x) > 0$) indicates that the worker is willing to accept less than her market value to benefit the employer, while a negative wage gap (hence, $RWG(x) < 0$) suggests the worker's reluctance to stay unless the wage exceeds her market value. 
However, the wage gap's ability to measure the worker's motivation may be distorted by the scaling effect. Specifically, if a worker consistently accepts a wage at a fixed proportion of her market value, the wage gap increases with $x$ solely due to the scaling of $v(x)$.\footnote{For example, suppose a worker always accepts a wage equal to 90\% of $v(x)$. At $x=1$, the wage gap is $v(1)-10\%v(1)=15$, while at $x=2$, it becomes $v(2)-10\%v(2)=25$.} To control for this effect, we normalize the wage gap by $v(x)$, ensuring that $RWG(x)$ reflects the fraction of the training return the worker is willing to forgo as compensation to the employer when the wage gap is positive.

Based on the classical model's prediction, the worker's $MAW(x)$ would equal her market value $v(x)$, leading to a relative wage gap ($RWG(x)$) of zero across all levels of $x$. This forms the basis for our first hypothesis:

\begin{hypothesis}\label{H1}
    In both \textit{ENDO} and \textit{EXO}, workers will set their minimum acceptable wages ($MAW(x)$) equal to their market wages ($v(x)$) at each training level, leading to a relative wage gap ($RWG(x)$) of zero at all training levels.
\end{hypothesis}

Our experiment also collects data on the discretionary effort that a worker is willing to contribute at each training level, $x \in \{0,1,2,3,4\}$.  The corresponding discretionary effort at level $x$, denoted by $DE(x)$, captures the worker's motivation to voluntarily benefit the employer at that level. While discretionary effort has been largely overlooked in the literature, the classical model can be extended to predict that a self-interested worker will choose $DE(x) = 0$, as contributing effort provides no financial benefit and incurs disutility.

\begin{hypothesis}\label{H2} 
In both \textit{ENDO} and \textit{EXO}, workers will choose zero discretionary effort, i.e., $DE(x)=0$, across all training levels, $x\in \{0,1,2,3,4\}$. 
\end{hypothesis}

\subsection{An Alternative Perspective: Predictions with Worker Reciprocity}\label{subsec_BE_model}

\noindent The primary hypotheses of the experiment are based on the classical model, which assumes that subjects' behavior is driven by self-interest. While these hypotheses provide a valuable benchmark, they may not fully capture the complexity of decision-making in workplace settings. Evidence suggests that, beyond self-interest, individuals often exhibit other-regarding preferences, particularly reciprocity, which can significantly shape workplace relationships.

Reciprocity has been widely studied and supported by experimental research. Laboratory studies (e.g., \cite{fehr93giftexchange, hannan2002lab, charness2011lab}) and field evidence (e.g., \cite{cohn2014fair}) suggest that reciprocal behavior matters and have long-term effects in the workplace. Exploring reciprocity allows us to examine its impact on worker behavior and compare it to our experimental findings on minimum acceptable wages and discretionary effort.

In this subsection, we assume that the employer, who may represent a company or institution, act selfishly by maximizing his own profits (monetary payoffs), while the worker is motivated by reciprocity. The worker's preferences are modeled by the utility function developed by \cite{DK2004}, an extension of \cite{rabin1993incorporating} to dynamic settings. This model accounts for sequential reciprocity, which, when applied to our training context, allows the worker's reciprocal motivations to vary by training level. We use a version of this model tailored to our experimental setup.

To analyze sequential reciprocity in the training context, it is important to recall the timeline of decisions outlined in Section \ref{sec: background}. In $T_0$, the employer chooses a training level $x \in \{0,1,2,3,4\}$, and the worker privately chooses a discretionary effort level ($DE$). In $T_1$, the employer makes a new offer and the worker decides whether to accept it. 

Let $s_i \in S_i$ represent the \emph{strategy} of player $i \in I = \{E, W\}$, and let $b_{i}$ denote player $i$'s (point) belief about the counterpart's strategy.
Recall that $m_i: \prod_{j \in I} S_i \to \mathbb{R}$ represents player $i$'s total monetary payoff, as defined in Equations \eqref{eq_E_total_payoff} and \eqref{eq_W_total_payoff}. Since the employer's payoff, $m_E$, includes a component determined by chance, resulting in either $y_H$ or $y_L$, we denote his expected monetary payoff as $\Bar{m}_E$, calculated as:
\begin{equation}\label{eq_E_expected_total_payoff}
    \Bar{m}_E=
    \begin{cases}
        150-w_1+ (100-20) x+800\cdot p(DE) &\text{ if the worker stays,} \\
        100-20 x +800 \cdot p(DE) &\text{ if the worker quits.}
    \end{cases}
\end{equation}

Reciprocity depends on the worker's \emph{kindness} toward the employer and her \emph{perceived kindness} from the employer.
The worker's \textbf{kindness} ($W$ toward $E$) is defined as the difference between the expected monetary payoff she intends to provide and an equitable payoff:
\begin{equation}
    \kappa_{WE}(s_{W},b_{W})=\Bar{m}_{E}(s_{W},b_{W}) -\Bar{m}^e_E(b_{W}),
\end{equation}
where $\Bar{m}^e_E(b_W)$, the \emph{equitable payoff to the employer}, is the midpoint between the maximum and minimum monetary payoffs the worker can provide the employer:\footnote{In \cite{DK2004}, the calculation of equitable payoff excludes weakly dominated strategies. See also the discussion by \cite{dufwenberg2018modelling}. This condition is omitted here as it does not affect equilibrium outcomes (defined in \ref{appendix BE_proofs}) or optimal behavior in our setup.}
\begin{equation*} 
    \Bar{m}^e_E(b_{W}) = \dfrac{1}{2} \cdot \left[ \max_{s_{W} \in S_{W}} m_E (s_{W}, b_{W}) + \min_{s_{W} \in S_{W}} m_E (s_{W}, b_{W}) \right]. 
\end{equation*}

If $\kappa_{WE}>0$ (i.e., $\Bar{m}_{E}(s_{W},b_{W})>\Bar{m}^e_E(b_{W})$), the worker is considered \emph{kind} toward the employer; otherwise, she is \emph{unkind}.
In addition to her own kindness toward the employer, the worker's reciprocity motivation also depends on the kindness she perceives from the employer. Let $c_{W}$ represent the worker's belief about $b_{E}$. The worker's \textbf{perceived kindness} is the difference between her perception of the expected monetary payoff the employer intends to provide and an equitable payoff:\footnote{Since the worker's monetary payoff is independent of chance moves, we do not need to refer to her expected payoff.}
\begin{equation}
    \lambda_{WEW} (b_{W},c_{W}) =m_{W}(b_{W}, c_{W})-m^e_W(c_{W}),
\end{equation}
where the \emph{equitable payoff to the worker} is given by:\footnote{For simplicity, we omit the non-dominated condition, which does not affect equilibrium analysis, as the minimum monetary payoff equals the minimum non-dominated payoff.}
\begin{equation*}
    \Bar{m}^e_W(c_W)=\dfrac{1}{2}\cdot \left[ \max_{b_W\in S_{E}} m_W (b_W, c_W)   +\min_{b_W\in S_{E}} m_W (b_W, c_W) \right].
\end{equation*}
As with $\kappa_{WE}$, $\lambda_{WEW}$ can be either positive or negative, indicating the worker's perception of kindness or unkindness from the employer.

We assume that $\lambda_{WEW}$ depends only on the worker's monetary payoff $m_W(\cdot,\cdot)$, excluding the disutility associated with the real-effort task.
\footnote{
This assumption is based on the idea that the worker is unlikely to incorporate the employer's conjectures about the unverifiable effort level into her perception of kindness. It also avoids the counterintuitive possibility that a worker might view a high training level as unkind simply because she assumes that the employer intends for her to choose a high $DE$ with significant disutility.} However, $\lambda_{WEW}$ does reflect the worker's monetary payoffs resulting from wage renegotiation, allowing us to distinguish between observable responses (i.e., wage renegotiation) and unobservable responses (i.e., discretionary effort) in our analysis.

The worker's utility from reciprocity is defined as the product of her kindness and perceived kindness:
\begin{equation}\label{eq:reciprocity_u}
    \lambda_{WEW} \left(b_{W},c_{W}\right) \cdot \kappa_{WE}\left(s_{W},b_{W}\right)= \left(m_W(b_{W}, c_{W})-m^e_W(c_{W})\right)\cdot \left(\Bar{m}_{E}(s_{W},b_{W}) -\Bar{m}^e_E(b_{W})\right).
\end{equation}
This equation reflects the sign-matching property of reciprocity, where the worker prefers mutual kindness (both terms positive) or mutual unkindness (both terms negative).

In our tailored model, decision utility varies across periods and is introduced backward from $T_1$ to $T_0$. In $T_1$, the worker takes action after receiving training at level $x$ and being offered a new wage $w_1(x)$. In our analysis, we use $w_1(x)$ instead of $w_1$ to indicate that the new wage is contingent on the training level $x$. The worker's decision utility from choosing to \emph{stay} or \emph{quit} is given by:

\noindent For \emph{stay}:
\begin{equation}\label{eq:u1_stay}
    \begin{aligned}
        u^1_{W,stay} &= m^1_{W,stay} +\eta\cdot \lambda_{WEW}(b_W,c_W)\cdot \kappa_{WE, stay}(s_W, b_W) \\\
        &= w_1(x) +\eta\cdot \lambda_{WEW}(b_W,c_W)\cdot \kappa_{WE, stay}(s_W, b_W) .
    \end{aligned}
\end{equation}
For \emph{quit}:
\begin{equation}\label{eq:u1_quit}
    \begin{aligned}
        u^1_{W,quit} &= m^1_{W,quit} +\eta\cdot \lambda_{WEW}(b_W,c_W)\cdot \kappa_{WE, quit}(s_W, b_W), \\
        &= v_1(x) +\eta\cdot \lambda_{WEW}(b_W,c_W)\cdot \kappa_{WE, quit}(s_W, b_W).
    \end{aligned}
\end{equation}
Here, $\kappa_{WE,a_W}(s_W, b_W) $ for $a_W\in \{stay, quit\}$ represents the worker's kindness when $s_W$ prescribes $a_W$ in $T_1$. Moreover, the worker's beliefs, $b_W$ and $c_W$, are updated based on her observation of the employer's choices, $(x, w_1(x))$, with conditional notation omitted for readability.

The worker chooses to \emph{stay} in $T_1$ if $u^1_{W,stay}\geq u^1_{W,quit}$, which is simplified to: 
\begin{equation}\label{con:stay}
w_1(x)-v_1(x) +\eta\cdot \lambda_{WEW}(b_W,c_W)\cdot \Delta\kappa_{WE}(b_W)
   \geq 0,
\end{equation}
where \begin{equation*}
    \Delta\kappa_{WE}(b_W):=\kappa_{WE, stay}(s_W, b_W)- \kappa_{WE, quit}(s^{\prime}_W, b_W)
\end{equation*} represents the \textbf{relative kindness} toward the employer when the worker stays rather than quits.

In $T_0$, the worker decides on a level of discretionary effort ($DE$) after receiving training at level $x$. Her decision utility from choosing $DE$ is given by:
\begin{equation}\label{eq:u0}
    u^0_{W,DE} = m^0_{W} +\eta \cdot \lambda_{WEW}(b_W,c_W)\cdot  \kappa_{WE,DE}(s_W,b_W) -k(DE)
\end{equation}
where the $T_0$ monetary payoff $m^0_{W}$ is independent of $DE$, $k(DE)$ represents the non-monetary disutility from the real-effort task at level $DE$, assumed to be strictly increasing and convex, ensuring a unique interior solution under the first-order condition, and $\kappa_{W,DE}$ denotes the worker's kindness toward the employer for a given $DE$.

The predictions are derived from equilibrium analysis, adhering to the solution concept by \cite{DK2004}, in which both the worker and the employer maximize their decision utilities based on correct beliefs. We assume a sufficiently large parameter $\eta$ to ensure reciprocity meaningfully influences the worker's behavior, consistent with the focus of this subsection. A formal definition of the equilibrium concept is provided in the Appendix. These predictions illustrate how worker reciprocity may influence her minimum acceptable wage ($MAW$) and her choice of discretionary effort ($DE$), without attempting to explain the data or rigorously test the model.

\subsubsection*{Reciprocal worker behavior in ENDO}
\noindent A reciprocal worker will choose to \text{stay} if the condition in \eqref{con:stay} holds. This condition is not directly affected by the $DE$ chosen in $T_0$, as it depends solely on the worker's relative kindness, $\Delta \kappa_{WE}$, which is unaffected by $DE$. This is intuitive: once $DE$ is chosen in $T_0$, it is sunk by $T_1$, and its contribution to $\kappa_{WE}$ remains the same regardless of whether the worker stays or quits.

The other key component in the condition is the perceived kindness, $\lambda_{WEW}=\left(m_W(b_{W}, c_{W})-m^e_W(c_{W})\right)$. To obtain tractable numerical predictions, we simplify the equitable payoff, $m^e_W(c_{W})$, by assuming it is the midpoint between 
(i) the highest total wage available in our experimental setup, and (ii) the lowest acceptable wage in equilibrium:\footnote{Strict adherence to \cite{DK2004} would require $m^e_W(c_{W})$ to depend on the magnitude of $\eta$, which is difficult to identify in our experimental setup, making the prediction intractable.} 
\begin{equation*}
    m^e_W(c_{W})= \frac{w_0+\overline{w_1}+ w_0+\underline{w_1}}{2}=\frac{50+600 + 50+50}{2}=375.
\end{equation*}

Our model predicts that $\lambda_{WEW}$ is positive for $x = 3$ and $4$ and negative for $x = 0, 1, 2$. The derivation of this result, which hinges on our simplifying assumption for $m^e_W(c_{W})$, is detailed in the appendix. While the model shows distinct behavior patterns for the worker when $x$ is below $2$ or above $3$, these points are not intended for precise empirical testing. The general trend is the focus: if a worker only trades off self-interest and reciprocity, $RWG(x)$ is predicted to be zero at lower values of $x$ and positive at higher ones.

\begin{obs}
    In \textit{ENDO}, a reciprocal worker would set her minimum acceptable wages ($MAW$s) equal to the market wages at $x = 0, 1, 2$, resulting in zero relative wage gaps ($RWG(x) = 0$). At $x = 3, 4$, she would set her minimum acceptable wages ($MAW$s) lower than the market wages, resulting in positive relative wage gaps ($RWG(x) > 0$).
\end{obs}

Given that $\lambda_{WEW}$ is negative for $x = 0, 1, 2$, and positive for $x = 3$ and $4$, a reciprocal worker will not choose positive $DE$ when $x$ is low, as this would benefit an unkind employer. However, when $x$ is high and the disutility $k(DE)$ is not too large, the worker may choose positive $DE$ to benefit a kind employer.

\begin{obs}
    In \textit{ENDO}, a reciprocal worker would choose zero $DE$ at $x=0,1,2$ and a positive amount of $DE$ at $x=3,4$. 
\end{obs}

\subsubsection*{Reciprocal worker behavior in EXO}

\noindent In the \textit{EXO} treatment, where $x$ is exogenously determined, the worker encounters five independent scenarios corresponding to different training levels, $x \in \{0, 1, 2, 3, 4\}$. The worker's perceived kindness $\lambda_{WEW}$ under each $x$ differs dramatically from the corresponding levels in \textit{ENDO}, highlighting the distinction between the two treatments in our model.

Our model predicts that the minimum acceptable wage ($MAW$) for each scenario will equal the corresponding market wage. This result follows from our simplifying assumption, which causes the worker to perceive unkindness ($\lambda_{WEW} < 0$) when offered the market wage. Nevertheless, she will still accept the wage because the employer's profit remains unchanged regardless of whether the worker stays or quits.\footnote{As previously noted, this payoff structure was designed to exclude any surplus unrelated to training.} Consequently, the relative kindness, $\Delta \kappa_{WE}$, in Condition \eqref{con:stay} is zero. Furthermore, the worker would reject any wage lower than the market wage to avoid providing kindness to an unkind employer, making the market wage her minimum acceptable wage.

\begin{obs}
     In \textit{EXO}, a reciprocal worker would set her minimum acceptable wages ($MAW$s) equal to the market wages at each training level, resulting in a zero relative wage gap ($RWG(x)=0$) across all training levels.
\end{obs}

Since in \textit{EXO}, the employer is always perceived as unkind when offering the equilibrium wage $w_1(x)$, the reciprocal worker is never motivated to reward the employer.

\begin{obs}
    In \textit{EXO}, a reciprocal worker would choose zero $DE$ at all training levels, $x$. 
\end{obs}

\section{Results}\label{sec:results}

\subsection{Preliminaries}
\noindent We find no significant differences between the \textit{EXO} and \textit{ENDO} treatments in terms of gender composition (Fisher's Exact (FE), $p=0.580$), Social Value Orientation (SVO) types (FE, $p=0.571$), positive reciprocity scores (Mann-Whitney (MW), $p=0.593$), and negative reciprocity scores (MW, $p=0.705$).\footnote{One subject in the \textit{ENDO} treatment was excluded due to consistently demanding exceptionally high wages across all training levels. The results remain robust to the inclusion of this subject.}\textsuperscript{,}~\footnote{
In the \textit{ENDO} (\textit{EXO}) treatment, approximately 44\% (50\%) of participants are male, and approximately 58\% (resp. 64\%) are classified as ``prosocial'', with the remainder categorized as ``individualistic'' according to their SVO scores. The mean ($\pm$ standard deviation) positive and negative reciprocity scores ( $\pm$ standard deviation) are $4.34$ ($\pm 0.53$) and $3.5$ ($\pm 0.87$) in the \textit{ENDO} treatment, and $4.34$ ($\pm 0.66$) and $3.45$ ($\pm 0.90$) in the \textit{EXO} treatment.} These observations suggest that the subject samples are broadly comparable between both treatments.

We use the $X_0,X_1,\dots, X_4$ to denote training levels $x=0,1,\dots,4$ , respectively, where $X_0$ represents no training. 
Summary statistics in Table \ref{tab:summary} show that employers frequently chose to sponsor training (at $X_2$, $X_3$, and $X_4$). The \textit{ENDO\_NEG} and \textit{EXO\_NEG} treatments will be introduced later.

The employer recovers the training cost \emph{immediately} if the worker accepts a sufficiently reduced wage. We define the employer's \textbf{break-even threshold} ($BT(x)=c(x)/v(x)$, for $x>0$) as the \emph{positive} relative wage gap required for the employer to fully offset the training cost. When the worker accepts a sufficiently low wage, increasing $RWG(x)$ (i.e., $\frac{v(x)-MAW(x)}{v(x)}$) above $BT(x)$, training becomes profitable for the employer.

\begin{table}[htbp]
    \footnotesize 
    \centering
    \begin{tabu} to 1\textwidth {X[l] X[l] X[l]X[l]X[l]}
        \toprule
       & &   \multicolumn{1}{l}{Employer Decisions} & \multicolumn{2}{l}{Worker Decisions}\\
    \cmidrule(lr){3-3} \cmidrule(lr){4-5}
   Training level&  Break-even Threshold & Sponsored Training & $DE$& $RWG$\\
   $x$ & $BT$& (\%)& Mean (S.D)& Mean (S.D)\\
    \midrule 
    \multicolumn{2}{l}{\textbf{Panel A.} ENDO}\\
     $X_0$ & NA& 0.08 & $\underset{(3.08)}{1.63}^{***}$ & $\underset{(1.93)}{ \text{-}0.80 }^{***}$ \\
     $X_1$ & 0.13& 0.08 & $\underset{(3.08)}{2.71}^{***}$ & $\underset{(0.50)}{\text{-}0.10}$ \\
     $X_2$ & 0.16& 0.24 & $\underset{(3.12)}{3.80}^{***}$ & $\underset{(0.25)}{0.00}$ \\
     $X_3$ & 0.17& 0.34 & $\underset{(3.52)}{5.20}^{***}$ & $\underset{(0.18)}{0.05}^{**}$ \\
     $X_4$ & 0.18& 0.25 & $\underset{(4.06)}{6.59}^{***}$ & $\underset{(0.16)}{0.07}^{***}$ \\
    \midrule 
    \multicolumn{2}{l}{\textbf{Panel B.} EXO}\\
     $X_0$ & NA& 0.17 & $\underset{(3.38)}{2.38}^{***}$ & $\underset{(1.48)}{\text{-}0.60}^{***}$ \\
     $X_1$ & 0.13& 0.29 & $\underset{(3.24)}{3.12}^{***}$ & $\underset{(0.42)}{\text{-}0.11}^{**}$ \\
     $X_2$ & 0.16& 0.10 & $\underset{(3.30)}{4.03}^{***}$ & $\underset{(0.23)}{\text{-}0.01}$ \\
     $X_3$ & 0.17& 0.26 & $\underset{(3.48)}{4.84}^{***}$ & $\underset{(0.18)}{0.03}$ \\
     $X_4$ & 0.18& 0.17 & $\underset{(3.84)}{5.64}^{***}$ & $\underset{(0.17)}{0.05}$ \\
    \midrule
    \multicolumn{2}{l}{\textbf{Panel C.} ENDO\_NEG}\\
     $X_0$ & NA& 0.19 & $\underset{(4.31)}{3.62}^{***}$ & $\underset{(1.07)}{\text{-}0.80}^{***}$ \\
     $X_1$ & 0.13& 0.09 & $\underset{(3.74)}{3.69}^{***}$ & $\underset{(0.24)}{{-}0.09}^{***}$ \\
     $X_2$ & 0.16& 0.29 & $\underset{(3.64)}{3.71}$ & $\underset{(0.19)}{0.03}$ \\
     $X_3$ & 0.17& 0.31 & $\underset{(3.33)}{3.22}^{***}$ & $\underset{(0.21)}{0.08}^{*}$ \\
     $X_4$ & 0.18& 0.12 & $\underset{(3.62)}{3.19}^{***}$ & $\underset{(0.23)}{0.12}^{***}$ \\
    \midrule
    \multicolumn{2}{l}{\textbf{Panel D.} EXO\_NEG}\\
     $X_0$ & NA& 0.16 & $\underset{(3.34)}{2.20}^{***}$ & $\underset{(1.42)}{\text{-}0.73}^{***}$ \\
     $X_1$ & 0.13& 0.18 & $\underset{(3.11)}{2.46}^{***}$ & $\underset{(0.40)}{\text{-}0.16}^{***}$ \\
     $X_2$ & 0.16& 0.29 & $\underset{(3.23)}{2.66}^{***}$ & $\underset{(0.22)}{\text{-}0.04}^{**}$ \\
     $X_3$ & 0.17& 0.18 & $\underset{(3.54)}{2.93}^{***}$ & $\underset{(0.18)}{0.02}$ \\
     $X_4$ &  0.18& 0.20 & $\underset{(3.84)}{3.14}^{***}$ & $\underset{(0.16)}{0.04}$ \\
    \bottomrule
    \end{tabu}
    \begin{minipage}{1\textwidth} 
        \footnotesize
        \textbf{Notes:} The training cost is fully recovered by the employer when the relative wage gap ($RWG$) reaches $0.13$, $0.16$, $0.17$, and $0.18$ for training levels $1$, $2$, $3$, and $4$, respectively. In the \textit{ENDO} (\textit{ENDO\_NEG}) treatment, the proportion of employers sponsoring a positive training level is $0.08$, $0.08$, $0.24$, $0.34$, and $0.25$ ($0.19$, $0.09$, $0.29$, $0.31$, and $0.12$), respectively. The relative wage gap ($RWG$) is computed for all training levels ($X_0$ to $X_4$) for each worker. Statistical significance is based on a Signrank test, indicating whether $RWG$ is significantly different from zero. Statistical significance levels: *** $p<0.01$, ** $p<0.05$, * $p<0.1$.
    \end{minipage}
  \caption{Summary Statistics}
  \label{tab:summary}
\end{table}

\subsection{Positive treatments: ENDO and EXO}
\subsubsection{Worker: Discretionary effort}
\noindent In the experiment, all but two workers in the \textit{ENDO} and \textit{EXO} treatments completed at least the number of tables they had reported as their discretionary effort ($DE$) levels in the real-effort task. This suggests that subjects were unlikely to over-report their discretionary effort levels.

\begin{figure}[htbp]
    \centering
    \includegraphics[width=1\linewidth]{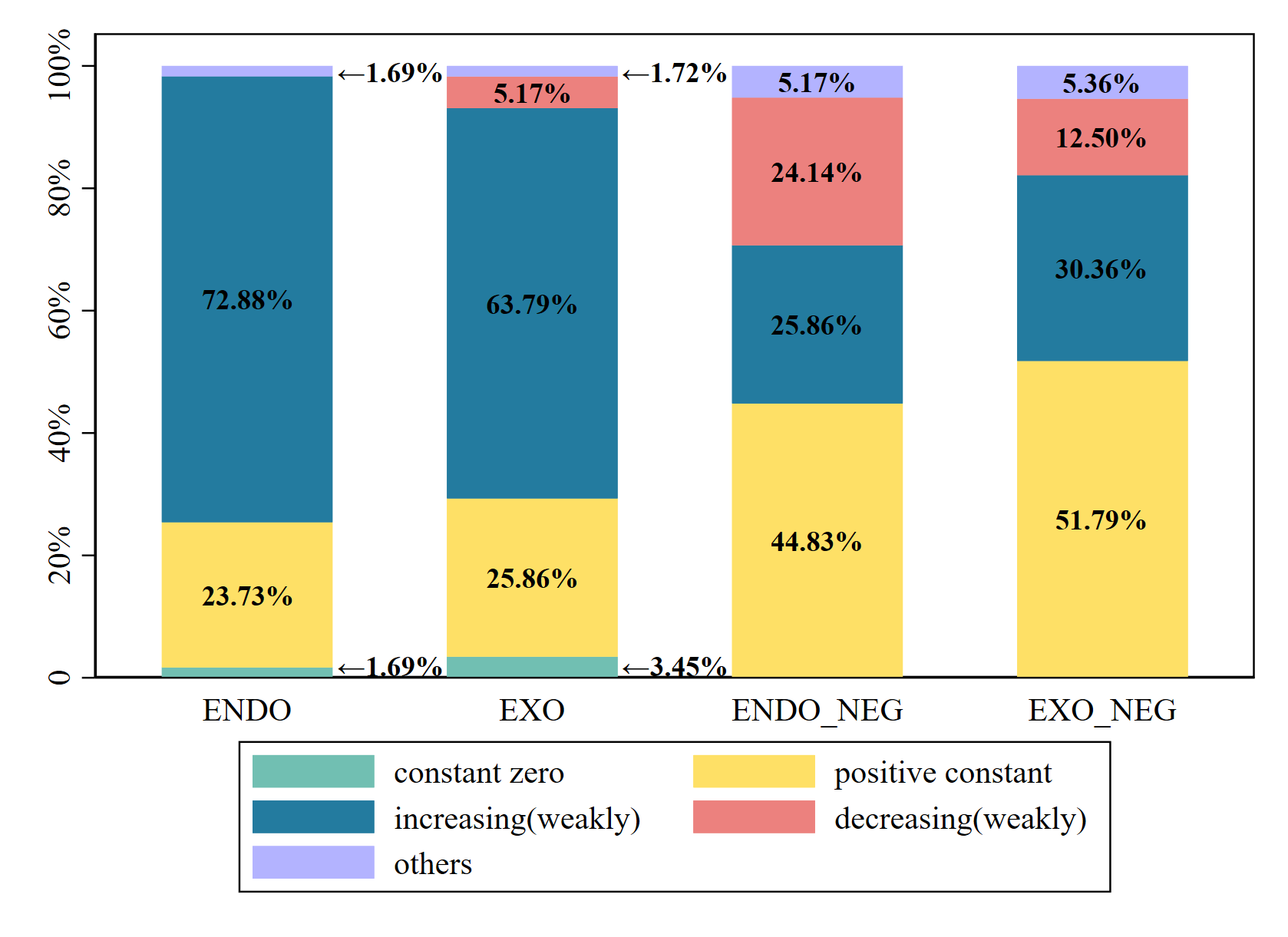} 
    \begin{minipage}{1\textwidth} 
        \footnotesize
        \textbf{Notes:} Subjects are categorized based on their effort provision. The patterns include: \emph{constant zero} (zero effort at all levels), \emph{positive constant} (consistent and positive $DE$), \emph{weakly-increasing} (at least one increase in $DE$), and \emph{weakly-decreasing} (at least one decrease in $DE$). Subjects who do not fit these patterns are categorized as \emph{others}.
    \end{minipage}
    \caption{Behavioral patterns of discretionary effort ($DE$) and counterproductive discretionary effort ($CDE$) across training levels ($x$).}
    \label{Figure_3}
\end{figure}

The summary statistics in Table \ref{tab:summary}, specifically from Panels A and B, show that workers' discretionary effort ($DE$) in the \textit{EXO} and \textit{ENDO} treatments is not only positive (Signrank $p < 0.001$ for each $x$), but also tends to increase with $x$. Specifically, \autoref{Figure_3} reveals that fewer than $4\%$ of workers exert zero discretionary effort at all training levels. In contrast, around $24\%$ of workers maintain a constant but positive level of $DE$ across all training levels. Finally, most workers ($72.88\%$ in the \textit{ENDO} treatment and $63.79\%$ in the \textit{EXO} treatment) exhibit weakly increasing $DE$, meaning their effort is non-constant but with at least one increase across training levels.

Building on these observed patterns, we estimate a multilevel linear model with observations nested within individuals. This approach accounts for the repeated-measures nature of the data, where each worker reports effort decisions across multiple training levels, ensuring proper control for within-subject dependencies. Table \ref{tab:econde} reports the results.

\begin{table}[htbp]
    \footnotesize 
    \centering
    \begin{tabu} to 1\textwidth {X[2l] X[l] X[l]}
        \toprule
        \toprule
        \multicolumn{3}{l}{\textbf{Dep. variable:} Discretionary Effort ($DE$) or Counterproductive Discretionary Effort ($CDE$)}\\
        \\
        & EXO \& ENDO& EXO\_NEG \& ENDO\_NEG\\
        \midrule 
        \multicolumn{3}{l}{\textit{Reference group}: $X_0$}\\
        ENDO & $\underset{(0.628)}{-0.752}$  & $\underset{(0.667)}{1.424}^{**}$  \\
        $X_1$ & $\underset{(0.294)}{0.741}^{**}$  & $\underset{(0.340)}{0.268}$  \\
        $X_2$ & $\underset{(0.294)}{1.655}^{***}$  & $\underset{(0.340)}{0.464}$  \\
        $X_3$ & $\underset{(0.294)}{2.466}^{***}$  & $\underset{(0.340)}{0.732}^{**}$  \\
        $X_4$ & $\underset{(0.294)}{3.259}^{***}$  & $\underset{(0.340)}{0.946}^{***}$  \\
        \midrule
     \multicolumn{3}{l}{\textit{Reference group}: EXO treatment}\\
        ENDO & $\underset{(0.628)}{-0.752}$  & $\underset{(0.667)}{1.424}^{**}$  \\
        ENDO $\times$ $X_1$ & $\underset{(0.413)}{0.343}$  & $\underset{(0.476)}{-0.199}$  \\
        ENDO $\times$ $X_2$ & $\underset{(0.413)}{0.514}$  & $\underset{(0.476)}{-0.378}$  \\
        ENDO $\times$ $X_3$ & $\underset{(0.413)}{1.111}^{***}$  & $\underset{(0.476)}{-1.129}^{**}$  \\
        ENDO $\times$ $X_4$ & $\underset{(0.413)}{1.707}^{***}$  & $\underset{(0.476)}{-1.377}^{***}$  \\
        \midrule
        Constant & $\underset{(0.446)}{2.379}^{***}$  & $\underset{(0.475)}{2.196}^{***}$  \\
        Observations & 585 & 570 \\
        $\chi^2$ test: $X_1=X_2$& $9.69^{***}$& 0.33\\
        $\chi^2$ test: $X_2=X_3$& $7.69^{***}$& 0.62\\
        $\chi^2$ test: $X_3=X_4$& $7.30^{***}$& 0.40\\
        \bottomrule
          \bottomrule
    \end{tabu}
    \begin{minipage}{1\textwidth} 
        \footnotesize
        \textbf{Notes:} The model is estimated using a multilevel linear regression with observations at level-1 nested within subjects at level-2. The reference groups are $X_0$ (no training) for training level comparisons and the \textit{EXO} treatment for treatment comparisons. The $\chi^2$ tests assess whether $DE$ varies across training increments ($X_1$ to $X_2$, $X_2$ to $X_3$, and $X_3$ to $X_4$) in both treatments. Statistical significance levels: *** $p<0.01$, ** $p<0.05$, * $p<0.1$.  
    \end{minipage}
  \caption{Multilevel linear regression estimates of the discretionary effort ($DE$) by training level}\label{tab:econde}
\end{table}

We use a multilevel linear model to study how workers' discretionary effort ($DE$) varies across training levels and treatments.
The estimates, reported in the first column of Table \ref{tab:econde}, can be summarized as follows.
Discretionary effort ($DE$) is significantly higher at any positive training level ($X_1, X_2, X_3, X_4$) compared to $X_0$ in both \textit{ENDO} and \textit{EXO} treatments ($p \leq 0.007$).
While $DE$ is generally higher in \textit{ENDO} than in \textit{EXO}, the difference is only significant for $X_3$ ($p<0.001$) and $X_4$ ($p<0.001$).
This brings us to our first two results.

\begin{result}\label{r1}
Contrary to Hypothesis 1, workers provide discretionary effort ($DE$) that benefits the employer, and they do so even when training is exogenously determined. Furthermore, workers' discretionary effort increases with the level of general training. This observation holds true irrespective of whether the level of general training is determined by the employer (\textit{ENDO}) or imposed exogenously (\textit{EXO}).
\end{result}

\begin{result}\label{r2}
    Workers' discretionary effort ($DE$) is significantly higher (at the $0.01$ level) at high training levels ($x=3$ or $x=4$) in the \textit{ENDO} treatment, where training is determined by the employer, compared to the \textit{EXO} treatment, where training is imposed exogenously.
\end{result}

Together, Results \ref{r1} and \ref{r2} suggest that while workers respond positively to employer-sponsored training by increasing their discretionary effort, their response is also shaped by their perception of the employer's training intentions. Specifically, workers exert greater discretionary effort when they know that the employer has chosen to sponsor a high level of training, even when the potential returns are uncertain.

\subsubsection{Workers: New wage demands}
\noindent We now examine another dimension of workers' decisions: the relative wage gap ($RWG$), defined as the difference between the market wage $v(x)$ and the minimum acceptable wage, scaled by $v(x)$. A positive $RWG$ indicates a \emph{wage discount}, where the worker demands a wage lower than $v(x)$, while a negative $RWG$ indicates a \emph{wage premium}.
According to the classical model, neither a wage discount nor a wage premium should be observed. This assumption underpins the conventional argument that providing training is not profitable.

The summary statistics from Panel A in Table \ref{tab:summary} show that in the \textit{ENDO} treatment, the relative wage gap ($RWG$) is significantly positive for $X_3$ (Signrank, $p = 0.026$) and $X_4$ (Signrank, $p = 0.001$). However, in both cases, $RWG$ remains significantly below the break-even threshold, suggesting that employers would be unable to fully recover training costs if workers were employed at their demanded wage. These findings align with the conventional argument that sponsoring general training provides no economic benefit to employers.  
On the other hand, in the \textit{EXO} treatment, $RWG$ is not significantly greater than zero for $X_3$ (Signrank, $p = 0.691$) or $X_4$ (Signrank, $p = 0.436$).

Furthermore, we observe that workers' $RWG$ is significantly lower than zero (Signrank, $p<0.001$ for the \textit{ENDO} and \textit{EXO} treatments) when the employer offers no training ($X_0$). This indicates that workers demanded wage premiums, suggesting their reluctance to stay with their employers.\footnote{This observation aligns with previous empirical evidence suggesting that providing general training reduces turnover rates \citep{cappelli2004whypaycollege,pattie2006tuition,manchester2008effect}.}

\begin{table}[htbp]
    \footnotesize 
    \centering
    \begin{tabu} to 1\textwidth {X[2l] X[l] X[l]}
        \toprule
        \toprule
        \multicolumn{3}{l}{\textbf{Dep. variable:} Relative Wage Gap ($RWG$)}\\
        \\
        & EXO \& ENDO& EXO\_NEG \& ENDO\_NEG\\
        \midrule 
\multicolumn{3}{l}{\textit{Reference group}: $X_0$}\\
$X_1$ & $\underset{(0.132)}{0.490}^{***}$  & $\underset{(0.104)}{0.566}^{***}$  \\
        $X_2$ & $\underset{(0.132)}{0.594}^{***}$  & $\underset{(0.104)}{0.685}^{***}$  \\
        $X_3$ & $\underset{(0.132)}{0.637}^{***}$  & $\underset{(0.104)}{0.746}^{***}$  \\
        $X_4$ & $\underset{(0.132)}{0.655}^{***}$  & $\underset{(0.104)}{0.769}^{***}$  \\
        \midrule 
        \multicolumn{3}{l}{\textit{Reference group}: EXO treatment}\\
ENDO & $\underset{(0.149)}{-0.195}$  & $\underset{(0.111)}{-0.074}$  \\
ENDO $\times$ $X_1$ & $\underset{(0.186)}{0.208}$  & $\underset{(0.146)}{0.144}$  \\
        ENDO $\times$ $X_2$ & $\underset{(0.186)}{0.206}$  & $\underset{(0.146)}{0.141}$  \\
        ENDO $\times$ $X_3$ & $\underset{(0.186)}{0.209}$  & $\underset{(0.146)}{0.132}$  \\
        ENDO $\times$ $X_4$ & $\underset{(0.186)}{0.219}$  & $\underset{(0.146)}{0.149}$  \\
        \midrule
        Constant & $\underset{(0.106)}{-0.604}^{***}$  & $\underset{(0.079)}{-0.725}^{***}$  \\
        Observations & 585 & 570 \\
         $\chi^2$ test: $X_1=X_2$& 0.61& 1.31\\
        $\chi^2$ test: $X_2=X_3$& 0.11& 0.35\\
        $\chi^2$ test: $X_3=X_4$& 0.02& 0.05\\
        \bottomrule
    \end{tabu}
    \begin{minipage}{1\textwidth} 
        \footnotesize
        \textbf{Notes:} The model is estimated using a multilevel linear regression with observations at level-1 nested within subjects at level-2. The reference groups are $X_0$ (no training) for training level comparisons and the \textit{EXO} treatment for treatment comparisons. The $\chi^2$ tests assess whether $RWG$ varies across training increments ($X_1$ to $X_2$, $X_2$ to $X_3$, and $X_3$ to $X_4$) in both treatments. Statistical significance levels: *** $p<0.01$, ** $p<0.05$, * $p<0.1$.
    \end{minipage}
  \caption{Multilevel linear regression estimates of the relative wage gap ($RWG$) by training level}\label{tab:econrwg}
\end{table}

Similar to the $DE$ analysis, we employ a multilevel linear model to examine the within- and between-treatment differences in $RWG$. The estimates are presented in Table \ref{tab:econrwg}. When training is sponsored ($x>0$), we find no significant effect of training levels on $RWG$ ($p\geq 0.253$ for each pairwise comparison). 
Moreover, we find no significant between-treatment differences in $RWG$ ($p\geq 0.306$) at any training level.

\begin{result}\label{r3}
     Contrary to Hypothesis 2, workers demanded significantly wage premiums ($RWG$ below zero) when no training was provided ($x=0$) in both the employer-determined training (\textit{ENDO}) and exogenously imposed training (\textit{EXO}) treatments. At higher training levels ($x=3$ and $x=4$), however, $RWG$ became significantly positive, indicating a shift toward wage discounts.  
     Nonetheless, across all training levels, $RWG$ remained well below the break-even threshold, suggesting that even with wage adjustments, employers were unable to fully recover training costs.
\end{result}

In addition, while wage renegotiation alone suggests no economic benefit from training investment for employers, incorporating discretionary effort ($DE$) provides a different perspective. Our experimental data show that when $DE$ is taken into account, employers' expected profits increase with training levels. Specifically, in the \textit{ENDO} treatment, expected profits rise with each increment in training: $326.57$ for $X_0$, 
$356.64$ for $X_1$, 
$389.02$ for $X_2$,
$435.86$ for $X_3$, 
and $484.45$ for $X_4$. In the \textit{EXO} treatment, expected profits also rise with training levels: $356.63$ for $X_0$, $371.28$ for $X_1$, $395.96$ for $X_2$, $419.55$ for $X_3$, and $441.55$ for $X_4$.

\subsection{Correlation between discretionary effort and relative wage gap}
\noindent
Finally, we examine the correlation between discretionary effort ($DE$) and the relative wage gap ($RWG$).

\begin{table}[htbp]
    \footnotesize 
    \centering
    \begin{tabu} to 1\textwidth {X[2l] X[l] X[l]X[l]X[l]X[l]}
        \toprule
        \toprule
        \multicolumn{6}{l}{\textbf{Dep. variable:} Relative Wage Gap ($RWG$)}\\
        \\
        & $X_0$& $X_1$& $X_2$& $X_3$& $X_4$\\
        \midrule 
  ENDO ($DE$)  & $\underset{(0.083)}{0.034}$  & $\underset{(0.021)}{0.022}$  & $\underset{(0.010)}{0.027}^{**}$  & $\underset{(0.006)}{0.015}^{**}$  & $\underset{(0.005)}{0.012}^{**}$  \\
  EXO ($DE$) & $\underset{(0.058)}{0.053}$  & $\underset{(0.017)}{0.024}$  & $\underset{(0.009)}{0.017}^{*}$  & $\underset{(0.007)}{0.011}$  & $\underset{(0.006)}{0.008}$  \\
 ENDO\_NEG ($CDE$) & $\underset{(0.033)}{-0.009}$  & $\underset{(0.009)}{-0.001}$  & $\underset{(0.007)}{0.004}$  & $\underset{(0.008)}{0.001}$  & $\underset{(0.008)}{0.006}$  \\
EXO\_NEG ($CDE$)  & $\underset{(0.051)}{-0.203}^{***}$  & $\underset{(0.016)}{-0.055}^{***}$  & $\underset{(0.009)}{-0.013}$  & $\underset{(0.007)}{0.005}$  & $\underset{(0.006)}{0.011}$  \\
\bottomrule
        \bottomrule
    \end{tabu}
    \begin{minipage}{1\textwidth} 
        \footnotesize
        \textbf{Notes:} The model is estimated using OLS, regressing the relative wage gap ($RWG$) on discretionary effort ($DE$) or counterproductive discretionary effort ($CDE$), with training levels ($X_0$ to $X_4$) as indicator variables. Statistical significance levels: *** $p<0.01$, ** $p<0.05$, * $p<0.1$.
    \end{minipage}
  \caption{OLS estimates of relative wage gap ($RWG$) on discretionary effort ($DE$) or counterproductive discretionary effort ($CDE$) across training levels}\label{tab:econrwgde}
\end{table}

As shown in Table \ref{tab:econrwgde}, no significant correlation is found in the \textit{EXO} treatment, whereas a positive correlation emerges in the \textit{ENDO} treatment for $x \geq 2$ ($p \leq 0.046$). This suggests that workers respond differently when training is employer-sponsored, aligning with our earlier findings that workers' responses are influenced by employers' training intentions.

Moreover, the findings across all subsections in this section remain robust when controlling for gender, SVO type, and reciprocity scores.

\section{Counterproductive Discretionary Effort}\label{sec:CDE}
\noindent The \textit{EXO} and \textit{ENDO} treatments show that workers willingly exert discretionary effort that benefits the employer. The intensity of effort exerted is influenced by both the level of sponsored training and whether it was provided voluntarily by the employer.
In the workplace, however, discretionary effort can take a negative form, where workers voluntarily engage in actions that reduce the firm's welfare, such as leaving negative reviews about their work experience. 
This raises the question of whether sponsored training influences workers' propensity to exert discretionary effort that is detrimental to employers. To examine this, we conducted two more treatments: \textit{ENDO\_NEG} (58 subjects, 3 sessions) and \textit{EXO\_NEG} (56 subjects, 3 sessions), conducted in late 2023.
These treatments mirror the previous treatments, except that the discretionary effort---now referred to as counterproductive discretionary effort ($CDE$)---reduces the probability that the employer will achieve the higher level of revenue, thereby reducing their employer's expected payoff.

\subsection{Negative treatments: ENDO\_NEG and EXO\_NEG}
\noindent In \textit{ENDO\_NEG} and \textit{EXO\_NEG} treatments, counterproductive discretionary effort ($CDE$), chosen by the worker, reduces the employer's likelihood of achieving the high long-term benefit, $y_H=800$, following the rule 
\begin{equation*}
    p(CDE)=0.8 -0.05 CDE \text{,}
\end{equation*}
where $p(CDE)$ ranges from $0.2$ to $0.8$, depending on the worker's chosen $CDE$ value in $\{0,\dots,12\}$.
The rest of the design follows the same structure, including that the total payoffs remain unchanged as specified in Equations \eqref{eq_E_total_payoff} and \eqref{eq_W_total_payoff}.
In the experiment, workers exert counterproductive discretionary effort by performing the real-effort task.
Finally, the two treatments differ in how sponsored training is determined: by the employer in \textit{ENDO\_NEG} or assigned randomly in \textit{EXO\_NEG}.

The classical model extends to the negative treatments, predicting that self-interested workers will not sacrifice their own monetary benefits for the employer.\footnote{An alternative perspective incorporates worker reciprocity, as specified in Section \ref{subsec_BE_model}. Within this framework, a reciprocal worker's $MAW(x)$ matches that in the corresponding positive treatments, \textit{ENDO} and \textit{EXO}. However, decisions regarding $CDE$ diverge significantly: In \textit{ENDO\_NEG}, $CDE(x)$ decreases with $x$ for $x = 0, 1, 2$ until it reaches zero, where it remains for $x = 3, 4$. In \textit{EXO\_NEG}, $CDE(x)$ decreases with $x$ until it reaches zero, but it may remain positive at $x = 3, 4$ if the disutility from completing the task, $k(CDE)$, is small. A detailed analysis is provided in \ref{appendix BE_proofs}. \label{ft:NEG_BE}} Accordingly:

\begin{hypothesis}\label{H3}
    In both \textit{ENDO\_NEG} and \textit{EXO\_NEG}, workers set their minimum acceptable wages ($MAW$) equal to their market wages at each training level, leading to a zero relative wage gap ($RWG(x)=0$) at all training levels.
\end{hypothesis}

Moreover, the classical model predicts that self-interested workers will choose zero $CDE$, since it yields no financial benefit and incurs disutility:

\begin{hypothesis}\label{H4} 
In both \textit{ENDO\_NEG} and \textit{EXO\_NEG}, workers choose zero counterproductive discretionary effort, i.e., $CDE(x)=0$, across all training levels, $x$. 
\end{hypothesis}

The summary statistics from Panels C and D in Table \ref{tab:summary} suggest that employer behavior on new wage offers is broadly similar across the \textit{ENDO\_NEG} and \textit{EXO\_NEG} treatments.\footnote{ differences in terms of gender composition (Fisher's Exact (FE), $p=0.581$), Social Value Orientation (SVO) types (FE, $p=0.851$), positive reciprocity scores (Mann-Whitney (MW), $p=0.369$), and negative reciprocity scores (MW, $p=0.844$).}
Turning to worker behavior, counterproductive discretionary effort ($CDE$) is significantly greater than zero in each comparison (Signrank,  $p < 0.010$).

To better understand workers' $CDE$ behavior across training levels, we refer to \autoref{Figure_3}. A substantial proportion of workers---$45\%$ in the \textit{ENDO\_NEG} treatment and $52\%$ in the \textit{EXO\_NEG} treatment --- maintain a consistently positive $CDE$ across all training levels. A notable share of workers exhibit weakly decreasing $CDE$ with $x$, though this proportion is considerably higher in \textit{ENDO\_NEG} ($24\%$) relative to \textit{EXO\_NEG} ($13\%$) treatment. Finally, some workers show a weakly increasing $CDE$ pattern in \textit{ENDO\_NEG} ($26\%$) and \textit{EXO\_NEG} ($30\%$) treatment, suggesting a greater tendency to diminish the employer's payoff as training levels rise. This behavior may stem from underlying preferences or unexpected motivations.

We use a multilevel linear model to analyze workers' counterproductive discretionary effort ($CDE$) across the \textit{EXO\_NEG} and \textit{ENDO\_NEG} treatments, with the results reported in Table \ref{tab:econde}. Pairwise comparisons between training levels (i.e., $X_0$ vs. $X_1$, $X_1$ vs. $X_2$, and so on) indicate that $CDE$ is not consistently influenced by training levels, suggesting that counterproductive discretionary effort is primarily driven by non-training-related factors, such as spite. 
Turning to between-treatment differences, we find that $CDE$ is significantly higher in \textit{ENDO\_NEG} at $X_0$ ($p=0.041$) but significantly lower in \textit{ENDO\_NEG} at $X_3$ ($p=0.018$) and $X_4$ ($p=0.004$). These results suggest that, while workers may actively reduce the employer's payoffs in response to the absence of training, they are less inclined to do so when higher levels of training are provided.

\begin{result}\label{r4}
While workers' counterproductive discretionary effort ($CDE$) remains strictly positive across all training levels in both treatments, it is significantly lower (at the $0.05$ level) at higher training levels ($x=3,4$) in the \textit{ENDO\_NEG} treatment, where training is employer-determined, compared to the \textit{EXO\_NEG} treatment, where training is exogenously assigned. Moreover, $CDE$ is significantly higher (at the $0.05$ level) at $x=0$ in the \textit{ENDO\_NEG} treatment.
\end{result}

Turning to wage renegotiation behavior, we observe that the relative wage gap ($RWG$) in the \textit{ENDO\_NEG} and \textit{EXO\_NEG} treatments mirrors the patterns in the positive treatments (\textit{ENDO} and \textit{EXO}), suggesting that the reversed impact of $CDE$ on the employer's payoff does not fundamentally alter workers' wage negotiation behavior. As in the positive treatments, workers are reluctant to stay at low levels of training, though this reluctance diminishes as training levels increase. Eventually, wage discounts emerge at higher training levels, but $RWG$ remains consistently well below the break-even threshold. This finding reaffirms the earlier observation that sponsoring general training yields no economic benefits for employers based on workers' wage renegotiation behavior alone.

\begin{result}\label{r5}
In line with the positive treatments, workers demanded significantly positive wage premiums ($RWG$ below zero) at lower training levels ($x=0$ and $x=1$) in both the employer-determined training (\textit{ENDO\_NEG}) and exogenously imposed training (\textit{EXO\_NEG}) treatments. At higher training levels ($x=3$ and $x=4$) in \textit{ENDO\_NEG}, however, $RWG$ became significantly positive, indicating a shift toward wage discounts. Nevertheless, across all training levels in both treatments, $RWG$ remained well below the break-even threshold, suggesting that even with wage adjustments, employers were unable to fully recover training costs.
\end{result}

Focusing solely on wage renegotiation might lead to the conclusion that sponsoring general training offers no economic benefit for employers. Incorporating the effects of workers' counterproductive discretionary effort ($CDE$) provides additional insight. Similar to our calculation in the \textit{ENDO} and \textit{EXO} treatments, we explore employers' expected profits by accounting for workers' punitive effort responses, using the experimental data. In contrast the positive treatments, employers' expected profits are not higher compared to the absence of training ($X_0$).\footnote{In the \textit{ENDO\_NEG} treatment, expected profits were $595.2$ at $X_0$, $578.26$ at $X_1$, $570.74$ at $X_2$, $588.96$ at $X_3$, and $594.64$ at $X_4$. In the \textit{EXO\_NEG} treatment, expected profits were $653.34$ at $X_0$, $625.8$ at $X_1$, $602.8$ at $X_2$, $584.14$ at $X_3$, and $564.85$ at $X_4$.}

Finally, we examine the relationship between counterproductive discretionary effort ($CDE$) and the relative wage gap ($RWG$). Table \ref{tab:econrwgde} shows no significant correlation in the \textit{ENDO\_NEG} treatment, while a significant negative correlation appears in \textit{EXO\_NEG} at $X_0$ and $X_1$ ($p\leq 0.01$). This suggests that the inverse impact of discretionary effort subtly influences workers' responses by affecting the relationship between $RWG$ and $CDE$.

\section{Conclusions and Discussions}\label{sec:conclusion}

\noindent Building on the rich literature exploring the economics of employer-sponsored general training, our study contributes to the understanding of why firms invest in general training despite the theoretical prediction of limited profitability due to post-training turnover and excessively high post-training wages \citep{becker,smithhayton1999drives,manchester2012reimbursement}. While much of the existing work focuses on wage-based explanations, this approach often falls short in fully accounting for general training investments, as many firms continue to sponsor such training even in the absence of clear wage compression benefits. This gap highlights the need to investigate alternative channels through which training affects worker behavior. Motivated by this need, we focus on a critical yet underexplored dimension: workers' effort-based responses. Specifically, we seek to answer the question: How do workers respond to employer-sponsored general training in terms of discretionary effort alongside wage renegotiation, and how are these responses influenced by employers' intentional provision of training?

Our experimental findings reveal that workers actively respond to employer-sponsored general training through both effort-based and wage-related behaviors. Workers' wage renegotiation patterns show reluctance to stay in the absence of training, as reflected in their demand for wage premiums at zero training levels. With increasing training, they adjust their willingness-to-accept in wage renegotiation, shifting toward wage discounts at higher training levels. However, these adjustments remain insufficient to cover the costs employers incur, consistent with the conventional view that sponsoring general training is unprofitable when evaluated solely through wage renegotiation,

Beyond monetary adjustments, our experimental evidence highlights the importance of workers' hidden effort responses. Workers voluntarily exert discretionary effort that benefits employers, deviating from the purely self-interested assumptions of classical models. This evidence underscores that this additional response is material and cannot be overlooked. Workers increase their discretionary effort with higher training levels, even when training is exogenously assigned. However, employer-determined training elicits a stronger effort-based response, reflecting workers' sensitivity to the intentional provision of training. These findings suggest that while training may not be directly profitable through wage compression alone, effort-based responses provide an additional channel through which employers can extract value from their investments.

Additionally, our exploration of punitive responses broadens the discussion of workplace reciprocation in laboratory experiments. While a few studies in the accounting area have begun to examine punitive behavior in controlled settings \citep{hannan2005bonusvpenalty,gonzalez2020penalty}, such behavior has received little attention within economics, despite being common in practice. For instance, employees dissatisfied with their workplace may provide negative feedback on review platforms. By incorporating these punitive actions into our analysis, we provide additional insights into firms' training investments and expand the experimental analysis of worker reciprocation in workplace settings from both positive and negative perspectives.

Our experimental observation reflects that, while punitive responses shift the nature of effort-based impacts, workers' wage renegotiation patterns remain similar to those observed in scenarios where effort benefits employers. Turning to effort-based responses, we find that, when accounting for a baseline level of spite, workers' punitive behavior is mitigated at higher training levels when the decision to invest is made by the employer, compared to when training is mandated, whereas it escalated in the absence of training. This comparative finding suggests that the role of employers' training intentions in shaping workers' punitive responses. By directly examining scenarios where workers can penalize employers through hidden actions, our study broadens the framework for analyzing the impact of firms' general training investments.

Lastly, our findings suggest that mandated training policies can be efficient, as increases in training levels, even when mandated, positively impact workers' motivation to reward their employers, extending the benefits of general training beyond merely enhancing workers' market value. However, our results also indicate that as training levels increase, the intensity of workers' effort to reward employers is weaker than when training is chosen by the employer, showing that the efficiency of mandated policies in enhancing benefits for employers could be limited. This highlights the need for further studies on how policy design can balance employer and worker incentives to maximize mutual benefits.

\clearpage

\bibliographystyle{chicago}
\bibliography{references}

\begin{thebibliography}{}

\bibitem[\protect\citeauthoryear{Acemoglu and Pischke}{Acemoglu and Pischke}{1998}]{acemoglu1998information}
Acemoglu, D. and J.-S. Pischke (1998).
\newblock Why do firms train? theory and evidence.
\newblock {\em Quarterly Journal of Economics\/}~{\em 113\/}(1), 79--119.

\bibitem[\protect\citeauthoryear{Acemoglu and Pischke}{Acemoglu and Pischke}{1999a}]{acemoglu1999beyondbecker}
Acemoglu, D. and J.-S. Pischke (1999a).
\newblock Beyond becker: Training in imperfect labour markets.
\newblock {\em Economic Journal\/}~{\em 109\/}(453), 112--142.

\bibitem[\protect\citeauthoryear{Acemoglu and Pischke}{Acemoglu and Pischke}{1999b}]{acemoglu1999structure}
Acemoglu, D. and J.-S. Pischke (1999b).
\newblock The structure of wages and investment in general training.
\newblock {\em Journal of Political Economy\/}~{\em 107\/}(3), 539--572.

\bibitem[\protect\citeauthoryear{Becker}{Becker}{1962}]{becker}
Becker, G.~S. (1962).
\newblock Investment in human capital: A theoretical analysis.
\newblock {\em Journal of Political Economy\/}~{\em 70\/}(5, Part 2), 9--49.

\bibitem[\protect\citeauthoryear{Bishop}{Bishop}{1997}]{bishop1996litreview}
Bishop, J.~H. (1997).
\newblock What we know about employer-provided training: A review of literature.
\newblock {\em Research in Labor Economics\/}~{\em 16}, 19--87.

\bibitem[\protect\citeauthoryear{Booth and Bryan}{Booth and Bryan}{2002}]{booth2002pays}
Booth, A.~L. and M.~L. Bryan (2002).
\newblock Who pays for general training? new evidence for british men and women.
\newblock {\em Working Paper\/}.

\bibitem[\protect\citeauthoryear{Brandts and Charness}{Brandts and Charness}{2011}]{brandts2011strategy}
Brandts, J. and G.~Charness (2011).
\newblock The strategy versus the direct-response method: a first survey of experimental comparisons.
\newblock {\em Experimental Economics\/}~{\em 14}, 375--398.

\bibitem[\protect\citeauthoryear{Cappelli}{Cappelli}{2004}]{cappelli2004whypaycollege}
Cappelli, P. (2004).
\newblock Why do employers pay for college?
\newblock {\em Journal of Econometrics\/}~{\em 121\/}(1-2), 213--241.

\bibitem[\protect\citeauthoryear{Charness}{Charness}{2004}]{charness2004attribution}
Charness, G. (2004).
\newblock Attribution and reciprocity in an experimental labor market.
\newblock {\em Journal of Labor Economics\/}~{\em 22\/}(3), 665--688.

\bibitem[\protect\citeauthoryear{Charness, Gneezy, and Henderson}{Charness et~al.}{2018}]{charness2018experimental}
Charness, G., U.~Gneezy, and A.~Henderson (2018).
\newblock Experimental methods: Measuring effort in economics experiments.
\newblock {\em Journal of Economic Behavior \& Organization\/}~{\em 149}, 74--87.

\bibitem[\protect\citeauthoryear{Charness and Kuhn}{Charness and Kuhn}{2011}]{charness2011lab}
Charness, G. and P.~Kuhn (2011).
\newblock Lab labor: What can labor economists learn from the lab?
\newblock In O.~Ashenfelter and D.~Card (Eds.), {\em Handbook of Labor Economics}, Volume~4, pp.\  229--331. Elsevier.

\bibitem[\protect\citeauthoryear{Charness, Masclet, and Villeval}{Charness et~al.}{2014}]{charness2014dark}
Charness, G., D.~Masclet, and M.~C. Villeval (2014).
\newblock The dark side of competition for status.
\newblock {\em Management Science\/}~{\em 60\/}(1), 38--55.

\bibitem[\protect\citeauthoryear{Cohn, Fehr, and Goette}{Cohn et~al.}{2014}]{cohn2014fair}
Cohn, A., E.~Fehr, and L.~Goette (2014).
\newblock Fair wages and effort provision: Combining evidence from a choice experiment and a field experiment.
\newblock {\em Management Science\/}~{\em 61\/}(8), 1777--1794.

\bibitem[\protect\citeauthoryear{Dohmen, Falk, Huffman, and Sunde}{Dohmen et~al.}{2009}]{dohmen2009surveywage}
Dohmen, T., A.~Falk, D.~Huffman, and U.~Sunde (2009).
\newblock Homo reciprocans: Survey evidence on behavioural outcomes.
\newblock {\em Economic Journal\/}~{\em 119\/}(536), 592--612.

\bibitem[\protect\citeauthoryear{Dufwenberg and Kirchsteiger}{Dufwenberg and Kirchsteiger}{2004}]{DK2004}
Dufwenberg, M. and G.~Kirchsteiger (2004).
\newblock A theory of sequential reciprocity.
\newblock {\em Games and Economic Behavior\/}~{\em 47\/}(2), 268--298.

\bibitem[\protect\citeauthoryear{Dufwenberg and Kirchsteiger}{Dufwenberg and Kirchsteiger}{2019}]{dufwenberg2018modelling}
Dufwenberg, M. and G.~Kirchsteiger (2019).
\newblock Modelling kindness.
\newblock {\em Journal of Economic Behavior \& Organization\/}~{\em 167}, 228--234.

\bibitem[\protect\citeauthoryear{Fehr, Kirchsteiger, and Riedl}{Fehr et~al.}{1993}]{fehr93giftexchange}
Fehr, E., G.~Kirchsteiger, and A.~Riedl (1993).
\newblock Does fairness prevent market clearing? an experimental investigation.
\newblock {\em Quarterly Journal of Economics\/}~{\em 108\/}(2), 437--459.

\bibitem[\protect\citeauthoryear{Gill and Prowse}{Gill and Prowse}{2011}]{gill2011novel}
Gill, D. and V.~L. Prowse (2011).
\newblock A novel computerized real effort task based on sliders.
\newblock {\em Working Paper\/}.

\bibitem[\protect\citeauthoryear{Gonzalez, Hoffman, and Moser}{Gonzalez et~al.}{2020}]{gonzalez2020penalty}
Gonzalez, G.~C., V.~B. Hoffman, and D.~V. Moser (2020).
\newblock Do effort differences between bonus and penalty contracts persist in labor markets?
\newblock {\em Accounting Review\/}~{\em 95\/}(3), 205--222.

\bibitem[\protect\citeauthoryear{Hannan, Hoffman, and Moser}{Hannan et~al.}{2005}]{hannan2005bonusvpenalty}
Hannan, R.~L., V.~B. Hoffman, and D.~V. Moser (2005).
\newblock Bonus versus penalty: does contract frame affect employee effort?
\newblock In R.~Zwick and A.~Rapoport (Eds.), {\em Experimental Business Research}, Volume~2, pp.\  151--169. Boston: Kluwer.

\bibitem[\protect\citeauthoryear{Hannan, Kagel, and Moser}{Hannan et~al.}{2002}]{hannan2002lab}
Hannan, R.~L., J.~H. Kagel, and D.~V. Moser (2002).
\newblock Partial gift exchange in an experimental labor market: Impact of subject population differences, productivity differences, and effort requests on behavior.
\newblock {\em Journal of Labor Economics\/}~{\em 20\/}(4), 923--951.

\bibitem[\protect\citeauthoryear{Kube, Mar{\'e}chal, and Puppe}{Kube et~al.}{2006}]{kube2006putting}
Kube, S., M.~A. Mar{\'e}chal, and C.~Puppe (2006).
\newblock Putting reciprocity to work-positive versus negative responses in the field.
\newblock {\em Working Paper\/}.

\bibitem[\protect\citeauthoryear{Kube, Mar{\'e}chal, and Puppe}{Kube et~al.}{2012}]{kube2012currency}
Kube, S., M.~A. Mar{\'e}chal, and C.~Puppe (2012).
\newblock The currency of reciprocity: Gift exchange in the workplace.
\newblock {\em American Economic Review\/}~{\em 102\/}(4), 1644--62.

\bibitem[\protect\citeauthoryear{Loewenstein and Spletzer}{Loewenstein and Spletzer}{1999}]{loewenstein1999general}
Loewenstein, M.~A. and J.~R. Spletzer (1999).
\newblock General and specific training: Evidence and implications.
\newblock {\em Journal of Human Resources\/}, 710--733.

\bibitem[\protect\citeauthoryear{Manchester}{Manchester}{2008}]{manchester2008effect}
Manchester, C.~F. (2008).
\newblock The effect of tuition reimbursement on turnover: A case study analysis.
\newblock In {\em The analysis of firms and employees: Quantitative and qualitative approaches}, pp.\  197--228. University of Chicago Press.

\bibitem[\protect\citeauthoryear{Manchester}{Manchester}{2012}]{manchester2012reimbursement}
Manchester, C.~F. (2012).
\newblock General human capital and employee mobility: How tuition reimbursement increases retention through sorting and participation.
\newblock {\em ILR Review\/}~{\em 65\/}(4), 951--974.

\bibitem[\protect\citeauthoryear{Montizaan, de~Grip, C{\"o}rvers, and Dohmen}{Montizaan et~al.}{2016}]{montizaan2016impact}
Montizaan, R., A.~de~Grip, F.~C{\"o}rvers, and T.~Dohmen (2016).
\newblock The impact of negatively reciprocal inclinations on worker behavior: Evidence from a retrenchment of pension rights.
\newblock {\em Management Science\/}~{\em 62\/}(3), 668--681.

\bibitem[\protect\citeauthoryear{Murphy, Ackermann, and Handgraaf}{Murphy et~al.}{2011}]{murphy2011measuring}
Murphy, R.~O., K.~A. Ackermann, and M.~J. Handgraaf (2011).
\newblock Measuring social value orientation.
\newblock {\em Judgment and Decision Making\/}~{\em 6\/}(8), 771--781.

\bibitem[\protect\citeauthoryear{Opitz and Schwaiger}{Opitz and Schwaiger}{2023}]{Timm}
Opitz, T. and C.~Schwaiger (2023).
\newblock Everyone likes to be liked: Experimental evidence from matching markets.
\newblock {\em Working Paper\/}.

\bibitem[\protect\citeauthoryear{Pattie, Benson, and Baruch}{Pattie et~al.}{2006}]{pattie2006tuition}
Pattie, M., G.~S. Benson, and Y.~Baruch (2006).
\newblock Tuition reimbursement, perceived organizational support, and turnover intention among graduate business school students.
\newblock {\em Human Resource Development Quarterly\/}~{\em 17\/}(4), 423--442.

\bibitem[\protect\citeauthoryear{Perugini, Gallucci, Presaghi, and Ercolani}{Perugini et~al.}{2003}]{perugini2003personal}
Perugini, M., M.~Gallucci, F.~Presaghi, and A.~P. Ercolani (2003).
\newblock The personal norm of reciprocity.
\newblock {\em European Journal of Personality\/}~{\em 17\/}(4), 251--283.

\bibitem[\protect\citeauthoryear{Rabin}{Rabin}{1993}]{rabin1993incorporating}
Rabin, M. (1993).
\newblock Incorporating fairness into game theory and economics.
\newblock {\em American Economic Review\/}, 1281--1302.

\bibitem[\protect\citeauthoryear{Sauermann}{Sauermann}{2023}]{sauermann2023effort}
Sauermann, J. (2023).
\newblock Worker reciprocity and the returns to training: Evidence from a field experiment.
\newblock {\em Journal of Economics \& Management Strategy\/}~{\em 32\/}(3), 543--557.

\bibitem[\protect\citeauthoryear{Smith and Hayton}{Smith and Hayton}{1999}]{smithhayton1999drives}
Smith, A. and G.~Hayton (1999).
\newblock What drives enterprise training? evidence from australia.
\newblock {\em International Journal of Human Resource Management\/}~{\em 10\/}(2), 251--272.

\bibitem[\protect\citeauthoryear{Xiao and Houser}{Xiao and Houser}{2005}]{xiaohouser2005emotion}
Xiao, E. and D.~Houser (2005).
\newblock Emotion expression in human punishment behavior.
\newblock {\em Proceedings of the National Academy of Sciences\/}~{\em 102\/}(20), 7398--7401.

\bibitem[\protect\citeauthoryear{Xiao and Houser}{Xiao and Houser}{2009}]{xiaohouser2009avoiding}
Xiao, E. and D.~Houser (2009).
\newblock Avoiding the sharp tongue: Anticipated written messages promote fair economic exchange.
\newblock {\em Journal of Economic Psychology\/}~{\em 30\/}(3), 393--404.

\end{thebibliography}

\newpage
\appendix
\renewcommand{\thesection}{Appendix \Alph{section}.}

\section{Proofs of the Reciprocity Model Predictions}
\label{appendix BE_proofs}

\noindent Our analysis adheres to the solution concept developed by \cite{DK2004}, known as the \emph{sequential reciprocity equilibrium} (referred to simply as ``the equilibrium'' where unambiguous), denoted by $\left(s^{\ast}_i, b^{\ast}_i, c^{\ast}_i\right)_{i\in \{E,W\}}$. This equilibrium based on the assumptions that 
\begin{itemize}
    \item The worker's strategy $s^{\ast}_W$ maximizes decision utilities, $u^t_W$, in each period $t$.
    \item The employer's strategy maximizes his monetary payoffs, $m_E$.
    \item All point beliefs are correct: $s^{\ast}_i=b^{\ast}_j=c^{\ast}_i$ for $\i\in \{E,W\}$
\end{itemize}

We assume that $\eta$ is sufficiently high for reciprocity to have a significant impact. Specifically, let $\eta> \frac{1}{25}$.

We now formally introduce the simplifying assumption used to derive tractable numerical predictions for the equitable payoff for the worker,
$m^e_W(c_{W})$. This equitable payoff plays a central role in the worker's perceived kindness, $\lambda_{WEW}=\left(m_W(b_{W}, c_{W})-m^e_W(c_{W})\right)$.

A strict adherence of the original model by \cite{DK2004} would not pin down the value of $m^e_W(c_{W})$, and consequentially $\lambda_{WEW}$. 
This indeterminacy arises because the maximum monetary payoff the employer could offer the worker depends on the highest wage the worker is willing to accept, which in turn hinges on the sensitivity parameter $\eta$. 
While a high wage is typically perceived as a kind act, a highly reciprocal worker (i.e., with high $\eta$) might hesitate to accept this wage, as doing so reduces the employer's profit. Thus, for certain ranges of $\eta$ the maximum monetary payoff depends on $\eta$, complicating the analysis.

To concentrate on how the incorporation of worker reciprocity affects the theoretical expectations for worker's behavior, we adopt the following simplifying assumption: 

\begin{quote}
    The maximum monetary payoff the employer can offer the worker is set at the upper bound of the new wage $w_1$ in our experiment, $600$. This assumption implies that the worker always accepts the maximum wage.
\end{quote}

However, no additional assumption is required for the minimum monetary payoff the employer can offer the worker, as it is uniquely determined in equilibrium. In our main treatment \textit{ENDO} (and \textit{ENDO\_NEG}), this value is $50$, achieved by choosing $x=0$ and $w_1(0)=50$.  As established in the main context, a market wage of $w_1(0)=v(0)=50$ is always acceptable because rejecting it would not reduce the employer's monetary payoff. To see why $50$ is the minimum acceptable wage in equilibrium, consider an alternative minimum acceptable wage $\Tilde{w}_1(0)<50$. Upon observing $x=0$ and $\Tilde{w}_1(0)$, the worker perceives unkindness ($\lambda_{WEW}<0$) as $\Tilde{w}_1(0)$ is defined to be less than the equitable payoff. By deviating to \emph{quit}, the worker would obtain monetary payoff $v(0)$, which exceeds $\Tilde{w}_1(0)$, and would also obtain a positive reciprocity utility, as $\Delta\kappa_{WE}<0$. This leads to a higher decision utility ($u_W^1$), contradicting the assumption of equilibrium. Thus, in equilibrium, $\min\limits_{s_W\in S_W }m^e_W(c_{W})=50+50=100$.\footnote{
Readers familiar with \cite{DK2004} may note that the equitable payoff is not defined based on ``efficient strategies'' that exclude wasteful moves (see the discussion by \cite{DK2004,dufwenberg2018modelling}). For simplicity, we omit this detail, as the restriction never bind in our context when focusing on equilibrium.} 

In the \textit{EXO} treatment (and \textit{EXO\_NEG}), this value is $v(x)=50+100x$ derived using a similar reasoning as above. The proof is left to the reader.

As discussed in the main context, this simplifying assumption has quantitative implications, particularly in determining which levels of 
$x$ are considered kind in equilibrium. However, it does not qualitatively alter the trends specified in the observations.

\paragraph{Proof for Observation 1:}
In equilibrium, the worker correctly anticipates the equilibrium wage $w^{\ast}_1(x)$ upon observing $x$. In the \textit{ENDO} treatment, condition \ref{con:stay}, under which the worker will \emph{stay}, is expressed as:

\begin{equation*}
    [w^{\ast}_1(x) - (50+100x)] +\eta \cdot [(50+100x)-w^{\ast}_1(x)]\cdot  \left[  50+w^{\ast}_1(x)-\frac{750}{2} \right] \geq 0
\end{equation*}
where $ 50+w^{\ast}_1(x)-\frac{750}{2} = 50+w^{\ast}_1(x) - \left[\frac{1}{2}\times (50+600) + \frac{1}{2}\times (50+50)\right]$ is $\lambda_{WEW}$ under the simplifying assumption.

Recall the argument in the main context that the market wage 
$v(x)$ is always acceptable. A profit-maximizing employer will then set
$w^{\ast}_1(x)=v(x)$ as long as $ 50+w^{\ast}_1(x)-\frac{750}{2} =w^{\ast}_1(x)-325<0$, since in this case, offering a kind wage is more costly than offering the market wage. This occurs when $x \leq 2$ as $50+100\times 2 < 325$.
On the other hand, for $x\geq 3$ not less than $3$, offering a kind wage that will be accepted by the reciprocal worker is less expensive than offering the market wage, leading to $w^{\ast}_1(x)=325 + \frac{1}{\eta}$, where we assumed that $\eta> \frac{1}{25}$.

This also implies that, in equilibrium, a reciprocal worker perceives unkindness ($\lambda_{WEW}<0$) when observing $x\in \{0,1,2\}$, and kindness ($\lambda_{WEW}>0$) when observing $x\in \{3,4\}$. 
While these predictions suggest distinct behavioral patterns for the worker based on whether $x$ is below $2$ or above $3$  the specific thresholds are shaped by the simplifying assumptions about equitable payoffs, as shown in the calculations. The key insight lies in the qualitative trend: $RWG(x)$ is predicted to be zero at lower values of $x$ and positive at higher values.

\paragraph{Proof for Observation 2:}

Recall that the worker's decision in $T_1$ regarding whether to \emph{stay} is independent of $DE$, as $DE$ is sunk by $T_1$. Consequently, when deciding $DE$, the worker does not need to account for its impact on her future decision in $T_1$. 
Thus, the worker chooses $DE^{\ast}$ that maximizes her decision utility function:

\begin{equation*}
    u^0_{W,DE} = m^0_{W} +\eta \cdot \lambda_{WEW}\cdot  \kappa_{WE,DE}(s_W,b_W) -k(DE),
\end{equation*}
where
\begin{equation*}
     \kappa_{WE,DE}(s_W,b_W)=(100 - 20x +200+800\times DE\times 5\% -m_W^e(\cdot) ).
\end{equation*}

Given that $k(\cdot)$ is increasing and convex. $DE^{\ast}$ satisfies the following first-order condition:
\begin{equation*}
\begin{cases}
    -k'(DE)+\eta \cdot  \lambda_{WEW}(\cdot) \cdot 40 =0 &\text{ if } \lambda_{WEW}>0,\\
    DE = 0 &\text{ if } \lambda_{WEW}<0. 
\end{cases}
\end{equation*}
Here, $\lambda_{WEW}$ is independent of $DE$. From the proof of Observation 1, we know that $\lambda_{WEW}>0$ when $x\geq 3$ and $\lambda_{WEW}<0$ when $x\leq 2$. Therefore, Observation 2 is confirmed.

\paragraph{Proof for Observation 3:}

In the \textit{EXO} treatment, condition \ref{con:stay}, given a training level $x\in \{0,1,2,3,4\}$, is expressed as:

\begin{equation*}
    [w^{\ast}_1(x) - (50+100x)] +\eta \cdot [(50+100x)-w^{\ast}_1(x)]\cdot  
     \left[  50+w^{\ast}_1(x)-\frac{50+600+50+v(x)}{2} \right]\geq 0
\end{equation*}

Observe that $w_1(x)=v(x)$, which is always acceptable, would never yield a positive value for $50+w^{\ast}_1(x)-\frac{50+600+50+v(x)}{2}$. Consequently, in equilibrium, the worker would never perceive kindness ($\lambda_{WEW}<0$). The observation follows.

\paragraph{Proof for Observation 4:}
 From the proof of Observation 3, we know that, in \textit{EXO}, $\lambda_{WEW}<0$ for all levels of $x$. Therefore, Observation 4 is confirmed.

\paragraph{Proof for results in negative treatments:}
In Section \ref{sec:CDE}, we briefly mentioned our predictions in footnote \ref{ft:NEG_BE}. Here, we provide the corresponding proofs. Since it has been established that the decision in $T_0$ direct effect on the on-going decision in $T_1$, the converse effect of $CDE$ would not alter the condition under which the worker would \emph{stay}. As a result, the worker's $RWG(x)$ follows the same patterns as those observed in the \textit{ENDO} or \textit{EXO} treatments, respectively.

Regarding $CDE$ in \textit{ENDO\_NEG}, the worker chooses $CDE$ to maximize: 
\begin{equation*}
    u^0_{W,CDE} = m^0_{W} +\eta \cdot \lambda_{WEW}\cdot  \kappa_{WE,CDE}(s_W,b_W) -k(CDE),
\end{equation*}
which is nearly identical to the decision utility in the \textit{ENDO} treatment, except that $\kappa_{WE,CDE}(s_W,b_W)$ now decreases in $CDE$ with expected payoffs of $5\% \times 800$ per increment of $CDE$.

The first-order condition is given by:
\begin{equation*}
\begin{cases}
    -k'(CDE)+\eta \cdot  \lambda_{WEW}(\cdot) \cdot 40 =0 &\text{ if } \lambda_{WEW}<0,\\
    CDE = 0 &\text{ if } \lambda_{WEW}>0. 
\end{cases}
\end{equation*}

From the proof of Observation 1, we know that $\lambda_{WEW}>0$ when $x\geq 3$ and $\lambda_{WEW}<0$ when $x\leq 2$, as shown in the proof of Observation 1. For $x\geq 3$, $CDE^{\ast}=0$. For $x\leq 2$, $CDE^{\ast}$ can be positive as long as $k(CDE^{\ast})$ is sufficiently small, and $CDE^{\ast}(x)$ decreases in $x$ within its positive segment.

Regarding $CDE$ in \textit{EXO\_NEG}, the decision utility in $T_0$ takes the same functional form as above. However, as established in Observation 3, the perceived kindness $\lambda_{WEW}<0$ for all levels of $x$. Consequently, $CDE^{\ast}(x)$ decreases in $x$ within its positive segment and may remain positive for higher levels of $x$ if $k(CDE)$ is sufficiently low.

\end{document}